\documentclass[twocolumn,usenatbib]{aastex62}

\usepackage{amsfonts,amssymb,amsbsy,amsmath}
\usepackage{mathtools}
\usepackage{upgreek}
\usepackage{natbib}
\usepackage{graphicx}
\usepackage{xifthen}
\usepackage{bigints}
\usepackage{booktabs}
\usepackage{nicefrac}
\usepackage{color}
\usepackage{url}
\usepackage{hyperref}
\usepackage[theorems]{tcolorbox}
\catcode`_=\active
\newcommand_[1]{\ensuremath{\sb{\mathrm{#1}}}}
\newcommand*{\logten}{\mathop{\log_{10}}}

\newcommand{\cpropto}[1]{\mathrel{\stackrel{\makebox[0pt]{\mbox{\normalfont\tiny #1}}}{\propto}}}
\newcommand{\newpar}{{}}

\newcommand{\pdfbig}[1]{{\pi\big(#1\big)}}

\newcommand{\up}{\operatorname}
\newcommand{\diff}{{\up{d}}}

\newcommand{\bs}{\boldsymbol}

\newcommand{\Mean}{{\bs{\up\mu}}}

\newcommand{\CovMat}{{\bs{\up\Sigma}}}

\newcommand{\numChar}{{n}}

\newcommand{\paraChar}{{p}}

\newcommand{\episChar}{{n}}

\newcommand{\npe}[1][]{{ \numChar_{ \ifthenelse{\isempty{#1}}{\paraChar\episChar}{{\paraChar\episChar,#1}} } }}

\newcommand{\epis}{{nois}}

\def\gtrsim{\mathrel{\hbox{\rlap{\hbox{\lower4pt\hbox{$\sim$}}}\hbox{$>$}}}}
\def\lessim{\mathrel{\hbox{\rlap{\hbox{\lower4pt\hbox{$\sim$}}}\hbox{$<$}}}}

\newcommand{\rmz}{{\rm z}}

\newcommand{\emodel}[1][]{{ M_{ \ifthenelse{\isempty{#1}}{\epis}{{\epis,#1}} } }}
\newcommand{\emodelz}[1][]{{ M^\rmz_{ \ifthenelse{\isempty{#1}}{\epis}{{\epis,#1}} } }}
\newcommand{\emodellgrb}[1][]{{ M^\lgrb_{ \ifthenelse{\isempty{#1}}{\epis}{{\epis,#1}} } }}

\newcommand{\observed}{{obs}}
\newcommand{\intrinsic}{{int}}
\newcommand{\eff}{{eff}}
\newcommand{\obsi}{{obs,i}}
\newcommand{\inti}{{int,i}}
\newcommand{\lgrb}{{\rm LGRB}}
\newcommand{\model}{{\mathcal{R}}}
\newcommand{\mint}{{\model_\intrinsic}}
\newcommand{\mobs}{{\model_\observed}}
\newcommand{\mintlgrb}{{\model_\intrinsic^\lgrb}}
\newcommand{\meff}{{\eta_\eff}}

\newcommand{\param}{{\bs{\theta}}}
\newcommand{\pobs}{{\param_\observed}}
\newcommand{\pint}{{\param_\intrinsic}}
\newcommand{\peff}{{\param_{eff}}}
\newcommand{\pintlgrb}{{\param_\intrinsic^\lgrb}}
\newcommand{\pz}{{\param_z}}

\newcommand{\eparam}[1][]{{ \param_{ \ifthenelse{\isempty{#1}}{\epis}{{\epis,#1}} } }}
\newcommand{\eparamz}[1][]{{ \bs\param^\rmz_{ \ifthenelse{\isempty{#1}}{\epis^\rmz}{{\epis,#1}} } }}
\newcommand{\eparamlgrb}[1][]{{ \bs\param^\lgrb_{ \ifthenelse{\isempty{#1}}{\epis^\lgrb}{{\epis,#1}} } }}

\newcommand{\domain}{{ \up{\Omega} }}
\newcommand{\domaindint}{{ \domain(\dint) }}

\newcommand{\domaindsetintlgrb}{{ \domain(\dsetintlgrb) }}
\newcommand{\domainpobs}{{ \domain(\pobs) }}

\newcommand{\data}{{\bs{D}}}
\newcommand{\dint}{{\data_\intrinsic}}
\newcommand{\dintp}{{\data^\possible_\intrinsic}}

\newcommand{\dinti}{{\data_\inti}}

\newcommand{\dobslgrb}{{\data_\observed^\lgrb}}
\newcommand{\dobsilgrb}{{\data_\obsi^\lgrb}}
\newcommand{\dobsilgrbmode}{{{\widehat\data}_\obsi^\lgrb}}
\newcommand{\dintlgrb}{{\data_\intrinsic^\lgrb}}
\newcommand{\dintilgrb}{{\data_\inti^\lgrb}}

\newcommand{\truth}{{\bs{R}}}
\newcommand{\possible}{{*}}
\newcommand{\truthset}{{\mathcal{R}}}

\newcommand{\truthsubset}[1][]{{ \truthset_{ \ifthenelse{\isempty{#1}}{\truth}{{\truth_{#1}}} } }}
\newcommand{\ptruthsubset}[1][]{{ \truthset_{ \ifthenelse{\isempty{#1}}{\truth}{{\truth_{#1}}} }^\possible }}

\newcommand{\dset}{{\mathcal{D}}}

\newcommand{\dsetobs}{{\dset_\observed^\lgrb}}
\newcommand{\dsetintlgrb}{{\dset_\intrinsic^\lgrb}}
\newcommand{\dsetobslgrb}{{\dset_\observed^\lgrb}}
\newcommand{\zset}{{\mathcal{Z}}}
\newcommand{\nint}{{N_{int}}}
\newcommand{\nobs}{{N_{obs}}}

\renewcommand*{\thefootnote}{\fnsymbol{footnote}}

\renewcommand*{\thefootnote}{\fnsymbol{footnote}}

\newcommand{\xx}[1][]{{ \ifthenelse{\isempty{#1}}{\textcolor{red}{XXX}}{\textcolor{red}{~(XXX {#1} XXX)~}} }}

\newcommand{\liso}{{L_{iso}}}
\newcommand{\eiso}{{E_{iso}}}
\newcommand{\epkz}{{E_{pz}}}
\newcommand{\durz}{{T_{90z}}}
\newcommand{\pbol}{{P_{bol}}}
\newcommand{\sbol}{{S_{bol}}}
\newcommand{\epk}{{E_{p}}}
\newcommand{\dur}{{T_{90}}}
\newcommand{\pph}{{P_{ph}}}
\newcommand{\zi}{{z_{i}}}
\newcommand{\lisoi}{{L_{iso,i}}}
\newcommand{\eisoi}{{E_{iso,i}}}
\newcommand{\epkzi}{{E_{pz,i}}}
\newcommand{\durzi}{{T_{90z,i}}}

\newcommand{\pbolimode}{{\widehat{P}_{bol,i}}}
\newcommand{\sbolimode}{{\widehat{S}_{bol,i}}}
\newcommand{\epkimode}{{\widehat{E}_{p,i}}}
\newcommand{\durimode}{{\widehat{T}_{90,i}}}

\newcommand{\ldis}{{d_L}}
\newcommand{\thresh}{{th}}
\newcommand{\threshM}{{\mu_\thresh}}
\newcommand{\threshS}{{\sigma_\thresh}}
\newcommand{\mz}{{\dot\zeta}}

\defcitealias{hopkins2006normalization}{H06}
\defcitealias{li2008star}{L08}
\defcitealias{butler2010cosmic}{B10}

\graphicspath{{./}{figures/}}

\shorttitle{BATSE LGRBs redshift catalog}
\shortauthors{Shahmoradi \& Nemiroff}

\begin{document}

\title{A Catalog of Redshift Estimates for 1366 BATSE Long-Duration Gamma-Ray Bursts:\\Evidence for Strong Selection Effects on the Phenomenological Prompt Gamma-Ray Correlations}

\correspondingauthor{Amir Shahmoradi, Robert Nemiroff}
\email{a.shahmoradi@uta.edu (AS), nemiroff@mtu.edu (RJN)}

\author{Amir Shahmoradi}
\affiliation{
Department of Physics, Data Science Program \\
The University of Texas at Arlington \\
Arlington, TX 76010, USA
}

\author{Robert J. Nemiroff}
\affiliation{
Department of Physics \\
Michigan Technological University \\
Houghton, MI 49931, USA
}

\begin{abstract}
    We present a catalog of the redshift estimates and probability distributions for 1366 individual Long-duration Gamma-Ray Bursts (LGRBs) detected by the Burst And Transient Source Experiment (BATSE). This result is based on a careful classification and modeling of the population distribution of BATSE LGRBs in the 5-dimensional space of redshift as well as intrinsic prompt gamma-ray emission properties: the isotropic 1024ms peak luminosity (\liso), the total isotropic emission (\eiso), the spectral peak energy (\epkz), as well as the intrinsic duration (\durz), while taking into account the complex detection mechanism of BATSE and sample incompleteness. The underlying assumption in our modeling approach is that LGRBs trace the Cosmic Star Formation Rate and that the joint 4-dimensional distribution of the aforementioned prompt gamma-ray emission properties follows a multivariate log-normal distribution.
    Our modeling approach enables us to constrain the redshifts of BATSE LGRBs to average uncertainty ranges of $0.7$ and $1.7$ at $50\%$ and $90\%$ confidence levels, respectively.
    We compare our predictions with the previous redshift estimates of BATSE GRBs based on the proposed phenomenological high-energy relations, including the lag-luminosity, the spectral peak energy-luminosity, and the variability-luminosity relations. Our predictions are almost entirely at odds with the previous estimates based on these phenomenological high-energy correlations, in particular with the estimates derived from the lag-luminosity and the variability-luminosity relations. There is, however, a weak but significant correlation of strength $\sim0.26$ between our predicted redshift estimates and those derived from the hardness-brightness relations. The discrepancies between the estimates can be explained by the strong influence of sample incompleteness in shaping the phenomenologically proposed high-energy correlations in the literature. The presented catalog here can be useful for demographic studies of LGRBs and studies of individual BATSE events.
\end{abstract}


\section{Introduction}
\label{sec:introduction}

Throughout almost a decade of continuous operation, the Burst And Transient Source Experiment (BATSE) onboard the now-defunct Compton Gamma-Ray Observatory \citep{meegan1992spatial} detected more than $2700$ Gamma-Ray Bursts (GRBs). The BATSE catalog of GRBs provided the first solid evidence for the existence of at least two classes of GRBs: the short-hard (SGRBs) and the long-soft (LGRBs) \citep[e.g.,][]{kouveliotou1993identification}.\newpar

Traditionally, new GRB events have been classified into one of the two classes based on a sharp cutoff on the bimodal distribution of the observed duration ($\dur$) of the prompt gamma-ray emission, generally set to $\dur\sim2-3[s]$. However, the dependence of the observed duration of GRBs on the gamma-ray energy and the detector's specifications \citep[e.g.,][]{fenimore1995gamma, nemiroff2000pulse, qin2012comprehensive} has prompted many studies in search of less-biased alternative methods of GRB classification, typically based on a combination of the prompt gamma-ray and afterglow emissions as well as the host galaxy's properties \citep[e.g.,][]{gehrels2009gamma, zhang2009discerning, shahmoradi2010hardness, goldstein2011new, shahmoradi2011possible, zhang2012revisiting, shahmoradi2013multivariate, shahmoradi2013gamma, shahmoradi2014classification, shahmoradi2015short, lu2014amplitude} or based on the prompt-emission spectral correlations in conjunction with the traditional method of classification \citep[e.g.,][]{qin2013statistical}.\newpar

Alternative hypotheses on the existence of more than two classes of GRBs with distinct progenitors have also been extensively discussed and considered, often based on the statistical analysis of the observed (as opposed to intrinsic) properties of GRBs \citep[e.g.,][]{horvath2008classification, mukherjee1998three, hakkila2001tools, balastegui2001reclassification, hakkila2004subgroups, horvath2006new, gehrels2006new, chattopadhyay2007statistical, virgili2009low, horvath2010detailed, levan2013new}, some of which has been debated and challenged \citep[e.g.,][]{hakkila2000fluence, hakkila2000gamma, hakkila2003sample, hakkila2004subgroups, hakkila2004dual, shahmoradi2013multivariate, zhang2014long, levan2013new}.\newpar

Ideally, the classification of GRBs should be independent of their cosmological distances from the earth and free from potential sample biases due to detector specifications, selection effects, sample incompleteness, and should solely rely on their intrinsic properties. Such classification methods are still missing in the GRB literature and hard to devise, mainly due to the lack of a homogenously-detected, sufficiently-large catalog of GRBs with measured redshifts.\newpar

A number of studies have already attempted to estimate the unknown redshifts of GRBs based on the apparently-strong phenomenological correlations observed between some of the spectral and temporal prompt gamma-ray emission properties of GRBs. The most prominent class of such relations are the apparently-strong correlations of the intrinsic brightness measures of the prompt gamma-ray emission (e.g., the total isotropic emission, $\eiso$, and the peak $1024ms$ luminosity, $\liso$) with other spectral or temporal properties of GRBs, such as {\it hardness} as measured by the intrinsic spectral peak energy $\epkz$ \citep[e.g.,][]{yonetoku2004gamma, yonetoku2014short}, light-curve variability \citep[e.g.,][]{fenimore2000redshifts, reichart2001possible}, the spectral lag \citep[e.g.,][]{schaefer2001redshifts}, or based on a combination of such relationships \citep[e.g.,][]{xiao2009estimating}.\newpar

These methods, however, can lead to incorrect or highly biased estimates of the unknown redshifts of GRBs if the observed high-energy correlations are constructed from a small sample of GRBs (typically the brightest events) with measured redshifts. Such small samples are often collected from multiple heterogeneous surveys and may neither represent the entire population of observed GRBs (with or without measured redshift) nor represent the unobserved cosmic population. More importantly, the potential effects of detector threshold and sample-incompleteness on them are poorly understood. Such biases manifest themselves in redshift estimates that are inconsistent with estimates from other methods, examples of which have been already reported by several authors \citep[e.g.,][]{guidorzi2005testing, ashcraft2007there, rizzuto2007testing, bernardini2014comparing}.\newpar

The selection effects in the detection, analysis, and redshift measurements of GRBs and their potential effects on the observed phenomenological high-energy correlations has been already extensively studied individually, in isolation from other correlations, \citep[e.g.,][]{hakkila2003sample, band2005testing, nakar2004outliers, butler2007complete, nava2008peak, shahmoradi2009real, butler2009generalized, butler2010cosmic, shahmoradi2011cosmological, shahmoradi2011possible, shahmoradi2013gamma, dainotti2015selection}. However, an ultimate resolution to the problem of estimating the unknown redshifts of GRBs in catalogs requires simultaneous multidimensional modeling of the intrinsic population distribution of GRB attributes, subject to the effects of detector threshold and sample incompleteness on their joint observed distribution \citep[e.g.,][]{butler2010cosmic, shahmoradi2013multivariate, shahmoradi2014classification, shahmoradi2015short}.\newpar

Building upon our previous studies in \citet{shahmoradi2013gamma,shahmoradi2013multivariate,shahmoradi2015short}, and motivated by the existing gap in the knowledge of the redshifts of LGRBs in BATSE catalog \citep[e.g.,][]{paciesas1999fourth, goldstein2013batse}, which as of today, constitutes the largest homogenously detected catalog of GRBs, here we present a methodology and modeling approach to constraining the redshifts of 1366 BATSE LGRBs. Despite lacking a complete knowledge of the true cosmic rate and redshift distribution of GRBs, we show that under the plausible assumption of LGRBs tracing the cosmic Start Formation Rate (SFR), we can still constrain the redshifts of individual BATSE LGRBs to within uncertainty ranges of width $0.7$ and $1.7$, on average, at $50\%$ and $90\%$ confidence levels, respectively. The presented work also paves the way towards a detector-independent minimally-biased phenomenological classification method for GRBs solely based on the intrinsic prompt gamma-ray data of individual events.\newpar

In the following sections, we present an attempt to further uncover some of the tremendous amounts of useful, yet unexplored information that is still buried in this seemingly archaic catalog of GRBs. Towards this, we devote \S\ref{sec:methods} of this manuscript to the development of redshift inference methodology, including brief discussions of the data collection procedure in \S\ref{sec:methods:data}, the proposed methodology for estimating redshifts in \S\ref{sec:methods:redshiftEstimation}, the cosmic SFR assumptions underlying our model in \S\ref{sec:methods:lgrbRateDensity}, the construction of an LGRB world model in \S\ref{sec:methods:lgrbWorldModel}, as well as a review of the BATSE LGRB detection algorithm and our approach to modeling sample incompleteness in \S\ref{sec:methods:modelingSampleIncompleteness}. The predictions of the model are provided in \S\ref{sec:results}, followed by a discussion of the implications of the results, comparison with previous independent redshift estimates and possible reasons for the observed discrepancies between the results of this study and the previous studies in \S\ref{sec:discussion}.

\section{Methods}
\label{sec:methods}

\subsection{BATSE LGRB Data}
\label{sec:methods:data}

    Fundamental to our inference problem is the issue of obtaining a dataset of BATSE LGRBs that is unbiased and representative of the population distribution of LGRBs detectable by BATSE Large Area Detectors (LADs). The traditional method of GRB classification based on a sharp cutoff line on the observed duration variable $\dur$ set at $2-3 [s]$ \citep[][]{kouveliotou1993identification} has been shown insufficient for an unbiased classification since the duration distributions of LGRBs and SGRBs have significant overlap \citep[e.g.,][]{butler2010cosmic, shahmoradi2013multivariate, shahmoradi2015short}. Instead, we follow the multivariate fuzzy classification approach of \citet{shahmoradi2013multivariate, shahmoradi2015short} to segregate the two populations of BATSE LGRBs and SGRBs based on their estimated observed spectral peak energies ($\epk$) from \citet{shahmoradi2010hardness} and $\dur$ from the current BATSE catalog \citep[][]{goldstein2013batse}. This leads us to a sample of 1366 BATSE LGRBs. We refer the interested reader to \citet{shahmoradi2013multivariate} for extensive details of the classification procedure. In addition to $\epk$ and $\dur$, we also collect the 1024 [ms] peak flux ($\pbol$) and the bolometric fluence (\sbol) of these events from the current BATSE catalog for inclusion in the analysis.\newpar

    Therefore, we represent the $i$th event, $\dobsilgrb$, in BATSE catalog by an a-priori `known' measurement uncertainty model, $\emodellgrb[i]$, that together with its `known' parameters, $\eparamlgrb[i]$, determine the joint 4-dimensional probability density function of the observed attributes of the event,
    \begin{eqnarray}
        \label{eq:dobsilgrb}
        \pi\big( \dobsilgrb &|& \emodellgrb[i] , \eparamlgrb[i] \big)  \nonumber \\
                            &\propto& \emodellgrb[i] \big( \dobsilgrb , \eparamlgrb[i] \big) ~.
    \end{eqnarray}

    The modes of these uncertainty models are essentially the values reported in the BATSE catalog and in \citet{shahmoradi2010hardness},
    \begin{equation}
        \label{eq:dobsilgrbmode}
        \dobsilgrbmode = [\pbolimode,\sbolimode,\epkimode,\durimode] ~,
    \end{equation}

    The entire BATSE dataset of 1366 LGRB events attributes is then represented by the collection of pairs of such uncertainty models and their parameters,
    \begin{equation}
        \label{eq:dsetobslgrb}
        \dsetobslgrb = \big\{ \big( \emodellgrb[i] , \eparamlgrb[i] \big) : ~1\leq i\leq1366 \big\} ~,
    \end{equation}

    Although, $\pbol$ as in \eqref{eq:dobsilgrbmode} is a conventionally-defined measure of the peak brightness and peripheral to and highly correlated with the more fundamental GRB attribute, $\sbol$, its inclusion in our GRB world model is essential as it determines, together with $\epk$, the peak {\it photon} flux, $\pph$, in $50-300$ [keV] range, based upon which BATSE LADs generally triggered on LGRBs.

\subsection{Redshift Estimation}
\label{sec:methods:redshiftEstimation}

    Our primary goal here is to constrain the probability density functions (PDF) of the redshifts of individual BATSE LGRBs, given the available observed BATSE dataset, $\dsetobs$. To achieve our goal, we model the process of LGRB observation as a non-homogeneous Poisson process whose mean rate parameter is the `observed' cosmic LGRB rate, $\mobs$. Thus, the probability of occurrence of each LGRB with intrinsic properties,
    \begin{eqnarray}
        && \dintilgrb = [\lisoi,\eisoi,\epkzi,\durzi] ~, \label{eq:dintlgrb} \\
        && \dinti = \{\dintilgrb,\zi :~1\leq i\leq 1366\} ~, \label{eq:dint}
    \end{eqnarray}

    \noindent in the 5-dimensional attributes space, $\domaindint$, of redshift (z), the 1024 [ms] isotropic peak luminosity ($\liso$), the total isotropic emission ($\eiso$), the intrinsic spectral peak energy ($\epkz$), and the intrinsic duration ($\durz$), as a function of the parameters, $\pobs$, of the observed LGRB rate model, $\mobs$, is given by,
    \begin{equation}
        \label{eq:modelGeneric}
        \pdfbig{ \dinti | \mobs , \pobs } \propto \mobs\big( \dinti , \pobs \big) ~,
    \end{equation}

    \noindent where $\mobs$ represents the BATSE-censored rate of LGRB occurrence in the universe. This can be further rewritten in terms of BATSE detection efficiency function, $\meff$, and the intrinsic cosmic LGRB rate, $\mint$, as,
    \begin{eqnarray}
        \label{eq:modelobs}
        \frac{\diff\nobs}{\diff\dint}
        &=& \mobs\big( \dint , \pobs \big) ~, \nonumber \\
        &=& \meff\big( \dint , \peff \big) \times \mint\big( \dint , \pint \big) ~,
    \end{eqnarray}

    \noindent for a given set of input intrinsic LGRB attributes, $\dint$, with $\pobs=\{\peff,\pint\}$ as the set of the parameters of our models for the BATSE detection efficiency and the intrinsic cosmic LGRB rate, respectively. Assuming, no systematic evolution of LGRB characteristics with redshift, which is a plausible assumption supported by independent studies \citep[e.g.,][]{butler2010cosmic} (hereafter: \citetalias{butler2010cosmic}), the intrinsic LGRB rate itself can be written as,
    \begin{eqnarray}
        \label{eq:modelint}
        \frac{\diff\nint}{\diff\dint}
        &=& \mint \big( \dint , \pint \big) \nonumber \\
        &=& \mintlgrb \big( \dintlgrb , \pintlgrb \big) 
        \times \frac{\mz(z,\pz)\nicefrac{\diff V}{\diff z}}{(1+z)}~,~~~~
    \end{eqnarray}

    \noindent with $\pint=\{\pintlgrb,\pz\}$, where $\mintlgrb$ is a statistical model with $\pintlgrb$ denoting its parameters, that describes the population distribution of LGRBs in the 4-dimensional attributes space of $\dint=[\liso,\eiso,\epkz,\durz]$, and the term $\mz(z,\pz)$ represents the comoving rate density model of LGRBs with the set of parameters $\pz$, while the factor $(1+z)$ in the denominator accounts for the cosmological time dilation. The comoving volume element per unit redshift, $\nicefrac{\diff V}{\diff z}$, is given by \citep[e.g.,][]{winberg1972gravitation, peebles1993principles},
    \begin{equation}
        \label{eq:dvdz}
        \frac{\diff V}{\diff z} = \frac{C}{H_0}\frac{4\pi {\ldis}^2(z)}{(1+z)^2\bigg[\Omega_M(1+z)^3+\Omega_\Lambda\bigg]^{1/2}} ~,
    \end{equation}

    \noindent with $\ldis$ standing for the luminosity distance,
    \begin{equation}
        \label{eq:ldis}
        \ldis(z)=\frac{C}{H_0}(1+z)\int^{z}_{0}dz'\bigg[(1+z')^{3}\Omega_{M}+\Omega_{\Lambda}\bigg]^{-1/2} ~.
    \end{equation}

    Throughout this work we assume a flat $\Lambda$CDM cosmology, with parameters set to $h=0.70$, $\Omega_M=0.27$ and $\Omega_\Lambda=0.73$ \citep{jarosik2011seven}. Also, the parameters $C$ \& $H_{0}=100h$ [km/s/MPc] stand for the speed of light and the Hubble constant respectively.\newpar

    If the three rate models, $(\mz,\meff,\mintlgrb)$, and their parameters were known a priori, one could readily compute the PDFs of the set of unknown redshifts of all BATSE LGRBs,
    \begin{equation}
        \label{eq:zset}
        \zset = \big\{ \zi: ~ 1\leq i\leq1366 \big\} ~,
    \end{equation}

    \noindent as,
    \begin{eqnarray}
        \label{eq:zPDF}
        \pi \big( \zset &|& \dsetobslgrb, \mobs, \pobs \big) \nonumber \\
        &\propto&   \int_{\domaindsetintlgrb} \mobs \big( \zset , \dsetintlgrb^* , \pobs \big) ~ \diff\dsetintlgrb^*,~~~~~~
    \end{eqnarray}

    \noindent where the integration is performed over all possible realizations, $\dsetintlgrb^*$, of the BATSE LGRB dataset given the measurement uncertainty models in \eqref{eq:dsetobslgrb}. The intrinsic properties, $\dintlgrb$, are exactly determined by the corresponding observed properties, $\dobslgrb$, via the following relations,
    \begin{align}
        \label{eq:obsIntMap}
        \liso &= 4\pi\times\ldis(z)^2\times\pbol &, \\
        \eiso &= 4\pi\times\ldis(z)^2\times\sbol / (z+1) &, \\
        \epkz &= \epk\times(z+1) &, \\
        \durz &= \dur / (z+1)^\alpha &,
    \end{align}

    \noindent with $\ldis(z)$ representing the luminosity distance as in \eqref{eq:ldis}, and $\alpha=0.66$ which takes into account the cosmological time-dilation as well as an energy-band correction (i.e., K-correction) of the form $(1+z)^{-0.34}$ to the observed durations \citep[e.g.][]{gehrels2006new}. For a range of possible parameter values, the redshift probabilities can be computed by marginalizing over the entire parameter space, $\domainpobs$, of the model,
    \begin{eqnarray}
        \label{eq:zMarginalPDF}
        \pi \big( \zset &|& \dsetobslgrb, \mobs \big) \nonumber \\
        &=& \int_\domainpobs \pdfbig{ \zset | \dsetobslgrb, \mobs, \pobs } \nonumber \\
        &\times& \pdfbig{ \pobs | \dsetobslgrb , \mobs } ~\diff\pobs ~.
    \end{eqnarray}

    The problem however, is that neither the rate models nor their parameters are known a priori. Even more problematic is the circular dependency of the posterior PDFs of $\zset$ and $\pobs$ on each other,
    \begin{eqnarray}
        \label{eq:paraPostProb}
        \pi \big( \pobs &|& \dsetobslgrb , \mobs \big) \nonumber \\
        &=& \int_{\domain(\zset)} \pdfbig{ \pobs \big| \zset , \dsetobslgrb , \mobs } \nonumber \\
        &\times& \pi \big( \zset \big| \dsetobslgrb , \mobs \big) ~\diff\zset ~.
    \end{eqnarray}

     Therefore, we adopt the following methodology, which is reminiscent of the Empirical Bayes \citep[][]{robbins1985empirical} and Expectation-Maximization algorithms \citep[][]{dempster1977maximum}, to estimate the redshifts of BATSE LGRBs. First, we propose models for $(\mz,\meff,\mintlgrb)$, whose parameters have yet to be constrained by observational data. Given the three rate models, we can proceed to constrain the free parameters of the observed cosmic LGRB rate, $\mobs$, based on BATSE LGRB data. The most appropriate fitting approach should take into account the observational uncertainties and any prior knowledge from independent sources. This can be achieved via Bayesian multilevel methodology \citep[e.g.,][]{shahmoradi2017multilevel} by constructing the likelihood function and the posterior PDF of the parameters of the model, while taking into account the uncertainties in observational data \citep[e.g., Eqn. 61 in][]{shahmoradi2017multilevel},
    \begin{widetext}
        \begin{eqnarray}
            && \pi \big( \pobs | \dsetobslgrb , \mobs \big) \nonumber \\
            &=& \frac
            {
                \pdfbig{ \pobs \big| \mobs }
            }{
                \pdfbig{ \dsetobslgrb \big| \mobs }
            }
            \int_{\domain(\zset)} \int_{\domaindsetintlgrb} \pdfbig{ \dsetintlgrb^* \big| \dsetobslgrb, \zset , \mobs , \pobs } \pdfbig { \zset \big| \mobs , \pobs } ~ \diff\dsetintlgrb^* ~ \diff\zset \label{eq:paraPostGeneric} \\
            &\cpropto{i.i.d.}&
            \exp \bigg( -\int_\domaindint \mobs \big( \dintp , \pobs \big) \diff\dintp \bigg)
            \prod_{i=1}^{1366}
            \meff \big( \dobsilgrbmode , \peff \big)
            \int_\domaindint \mint \big( \dintp , \pint \big) \pdfbig{ \dintp ~|~ \dobsilgrb } \diff\dintp , ~~~~~~ \label{eq:paraPostPoisson}
        \end{eqnarray}
    \end{widetext}

    \noindent where \eqref{eq:paraPostPoisson} holds under the assumption of independent and identical distribution (i.e., the i.i.d. property) of BATSE LGRBs, and the second integration within it is performed over all possible realizations of the truth for the $i$th observed BATSE LGRB, $\dobsilgrb$.\newpar

    Once, the posterior PDF of the model parameters is obtained, it can be plugged into \eqref{eq:zPDF} to constrain the redshift PDF of individual BATSE LGRBs at the second level of modeling.

\subsection{LGRB Redshift Distribution}
\label{sec:methods:lgrbRateDensity}

    Our main assumption in this work is that, the intrinsic comoving rate density of LGRBs closely traces the comoving Star Formation Rate (SFR) density, or a metallicity-corrected SFR density as prescribed by \citetalias{butler2010cosmic}. Consequently, regardless of the individually-unknown redshifts of BATSE LGRBs, the overall redshift distribution of all BATSE LGRBs together is enforced in our modeling to follow the cosmic SFR convolved with BATSE detection efficiency, $\meff$, which is modeled in \S\ref{sec:methods:modelingSampleIncompleteness}. This assumption is essential for the success of our modeling approach, as any attempts to constrain the comoving rate density, $\mz(z)$, of LGRBs solely based on BATSE data leads to highly degenerate parameter space, $\domainpobs$, and parameter estimates for our model, $\mobs$.\newpar

    As for the choice of the LGRB rate density model, $\mz$, we consider three different LGRB rate density scenarios, all three of which have the generic continuous piecewise form,
    \begin{equation}
        \label{eq:mz}
        \mz(z) \propto
        \begin{cases}
            (1+z)^{\gamma_0} & z<z_0 \\
            (1+z)^{\gamma_1} & z_0<z<z_1 \\
            (1+z)^{\gamma_2} & z>z_1 ~, \\
        \end{cases}
    \end{equation}

    \noindent with parameters,
    \begin{eqnarray}
        \label{eq:pz}
        \pz
        &=& (z_0,z_1,\gamma_0,\gamma_1,\gamma_2) \nonumber \\
        &=&
        \begin{cases}
            (0.97,4.5,3.4,-0.3,-7.8)    & \text{(\citetalias{hopkins2006normalization})} \\
            (0.993,3.8,3.3,0.055,-4.46) & \text{(\citetalias{li2008star})} \\
            (0.97,4.00,3.14,1.36,-2.92) & \text{(\citetalias{butler2010cosmic})}
        \end{cases}
    \end{eqnarray}

    \noindent corresponding to the SFR of \citet{hopkins2006normalization} (hereafter: \citetalias{hopkins2006normalization}), \citet{li2008star} (hereafter: \citetalias{li2008star}), and a bias-corrected redshift distribution of LGRBs derived from Swift data by \citetalias{butler2010cosmic}.

\subsection{LGRB Attributes Distribution}
\label{sec:methods:lgrbWorldModel}

    We model the joint 4-dimensional distribution of $\dintlgrb$, with a multivariate log-normal distribution, $\mintlgrb\equiv\mathcal{LN}$, whose parameters (i.e., the mean vector and the covariance matrix), $\pintlgrb=\{\bs\mu,\bs\Sigma\}$, will have to be constrained by data. The justification for the choice of a multivariate log-normal as the underlying intrinsic population distribution of LGRBs is multi-folded. First, the observed joint distribution of BATSE LGRB attributes highly resembles a log-normal shape that is censored close to BATSE's detection threshold. Second, unlike power-law, log-normal models provide natural upper and lower bounds on the total energy budget and luminosity of LGRBs, eliminating the need for setting artificial sharp bounds on the distributions in order to properly normalize them. Third, log-normal along with Gaussian distribution, are among the most naturally-occurring statistical distributions in nature, whose generalizations to multi-dimensions is also very well studied and understood. This is a highly desired property specially for our work, given the overall mathematical and computational complexity of the model proposed and developed here.

\subsection{Sample Incompleteness}
\label{sec:methods:modelingSampleIncompleteness}

    Compared to Fermi Gamma-Ray Burst Monitor \citep[][]{meegan2009fermi} and Neil Gehrels Swift Observatory \citep[][]{gehrels2004swift, lien2016third}, BATSE had a relatively simple triggering algorithm. The BATSE detection efficiency and algorithm has been already extensively studied by the BATSE team as well as independent authors \citep[e.g.,][]{pendleton1998batse, pendleton1995detector, hakkila2003sample} \citep[c.f.,][for further discussion and references]{shahmoradi2010hardness, shahmoradi2011possible, shahmoradi2013multivariate, shahmoradi2015short}. However, a simple implementation and usage of the known BATSE trigger threshold for modeling BATSE catalog's sample incompleteness can lead to systematic biases in the inferred quantities of interest. BATSE triggered on $>2700$ GRBs, out of which only $2145$ or approximately $79\%$ have been consistently analyzed and reported in the current BATSE catalog, with the remaining $21\%$ either having a low accumulation of count rates or missing a full spectral/temporal coverage \citep[][]{goldstein2013batse}. Thus, the extent of sample incompleteness in BATSE catalog may not be fully and accurately modeled by using the BATSE triggering algorithm alone.\newpar

\begin{figure}
        \centering
        \includegraphics[width=0.45\textwidth]{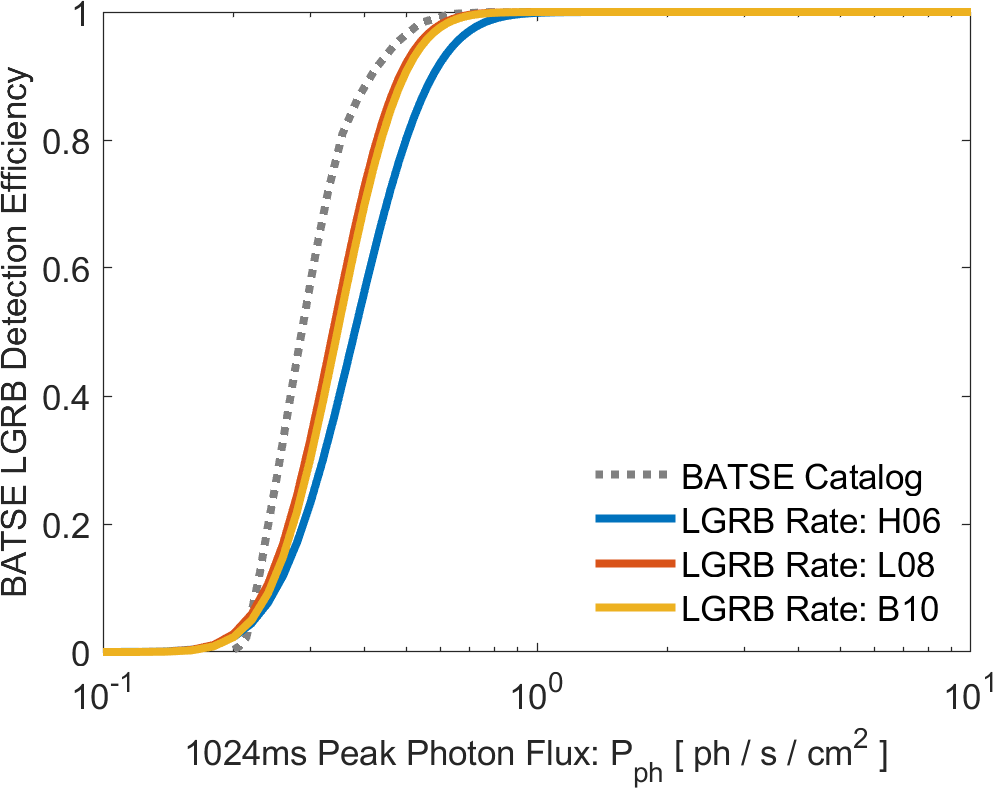}
        \caption{A comparison of the BATSE 4B Catalog nominal LGRB detection efficiency as a function of the 1024 [ms] peak photon flux, $\pph$, with the predicted detection efficiencies in this work based on the three different LGRB rate models considered: \citetalias{hopkins2006normalization}, \citetalias{li2008star}, \citetalias{butler2010cosmic}. The peak photon flux, $\pph$, is measured in the BATSE energy window $50-300$ [keV].\label{fig:BatseDetEff}}
    \end{figure}

    BATSE LADs generally triggered on a GRB if the number of photons per 1024 [ms] arriving at the detectors in $50-300$ [keV] energy window, \pph, reached above a certain threshold in units of the background photon count fluctuations, $\sigma$. This threshold was typically set to $5.5\sigma$ during much of BATSE's operational lifetime. However, the naturally-occurring fluctuations in the average background photon counts effectively lead to a monotonically increasing BATSE detection efficiency as a function of $\pph$, instead of a sharp cutoff on the observed $\pph$ distribution of LGRBs. Therefore, we model the effects of BATSE detection efficiency and sample incompleteness more appropriately by an Error function,
    \begin{eqnarray}
        \label{eq:methods:redshiftEstimation:efficiency}
        \pi\big(\mathrm{detection}&|&\threshM,\threshS,\pph\big) \nonumber \\
        &=& \frac{1}{2} + \frac{1}{2} \times \mathrm{erf}\bigg(\frac{\logten\pph-\threshM}{\threshS\sqrt{2}}\bigg) ~,
    \end{eqnarray}

    \noindent which gives the probability of the detection of an LGRB with $\pph$ for a given set of Error function's parameters, $\peff=\{\threshM,\threshS\}$. Due to the unknown effects of sample incompleteness on BATSE data, we leave these two threshold parameters free to be constrained by the observational data. Indeed, we show in Figure \ref{fig:BatseDetEff} that the resulting parameters of our BATSE detection efficiency model in \eqref{eq:methods:redshiftEstimation:efficiency} for all the three LGRB rates considered here indicate an effectively higher detection threshold than the nominal values reported in the BATSE catalog\footnote{Available at: \url{https://gammaray.nsstc.nasa.gov/batse/grb/catalog/4b/4br_efficiency.html}} \citep[see also][]{pendleton1995detector, pendleton1998batse, paciesas1999fourth, hakkila2003sample, shahmoradi2013multivariate, shahmoradi2015short}.\newpar

    The current BATSE catalog already provides estimates of $\pph$ for all 1366 LGRBs in our analysis. The connection between $\pph$ and the bolometric 1024 [ms] peak flux, $\pbol$, which is used in our modeling, is provided by fitting all LGRB spectra with a smoothly-broken power-law known as the Band model \citep[][]{band1993batse} to infer their spectral normalization constants, while fixing the low- and high-energy photon indices of the Band model to the corresponding population averages $(\alpha, \beta) = (-1.1,-2.3)$ and fixing the observed spectral peak energies, $\epk$, of individual bursts to the corresponding best-fit values from \citet{shahmoradi2010hardness}. Such an approximation is reasonable given the typically large uncertainties that exist in the spectral and temporal properties of GRBs \citep[e.g.,][]{butler2010cosmic, shahmoradi2013multivariate} and the relatively large variance of the population distribution of $\pbol$ in our sample.

\section{Results}
\label{sec:results}

    \begin{figure*}[tphb]
        \centering
        \begin{tabular}{cccc}
            \includegraphics[width=0.233\textwidth]{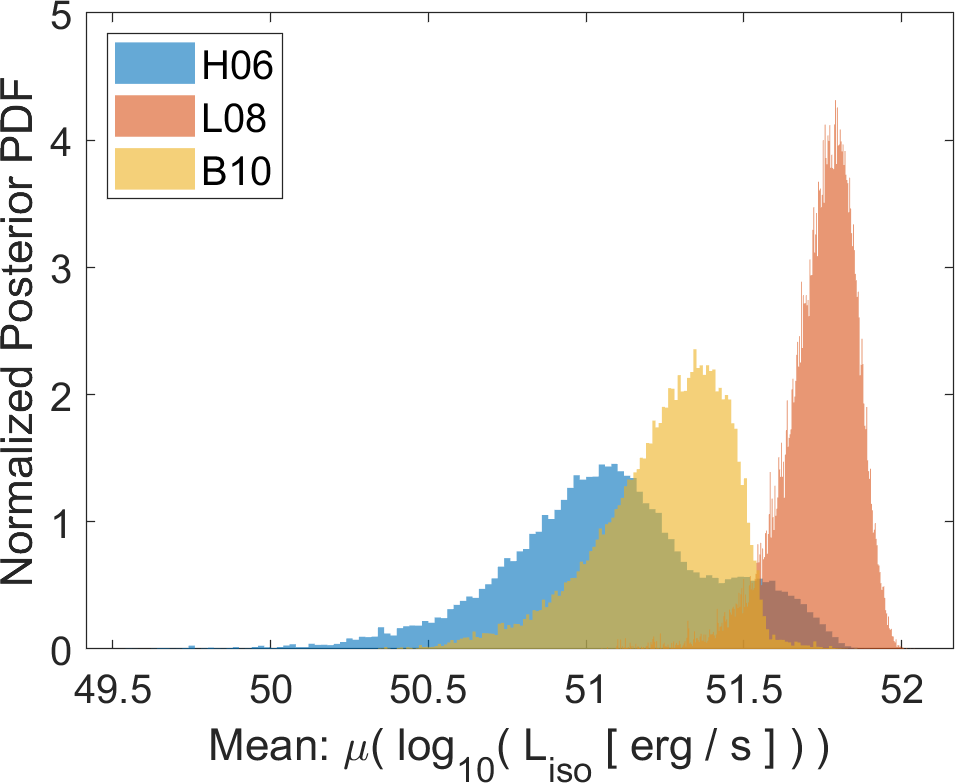} &
            \includegraphics[width=0.233\textwidth]{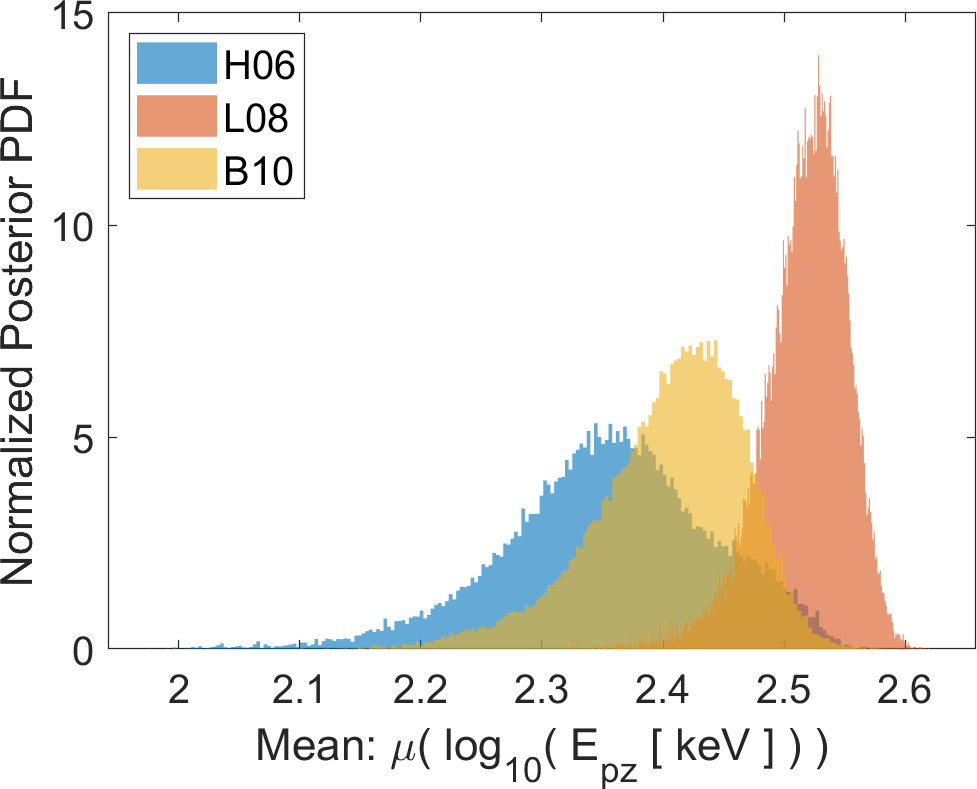} &
            \includegraphics[width=0.233\textwidth]{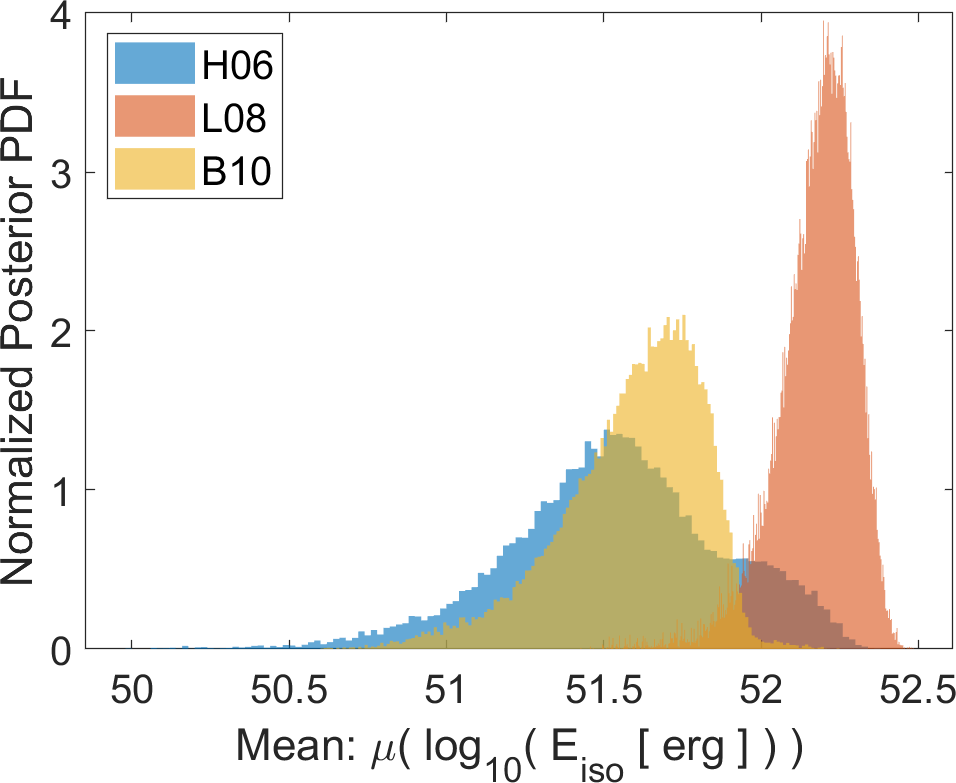} &
            \includegraphics[width=0.233\textwidth]{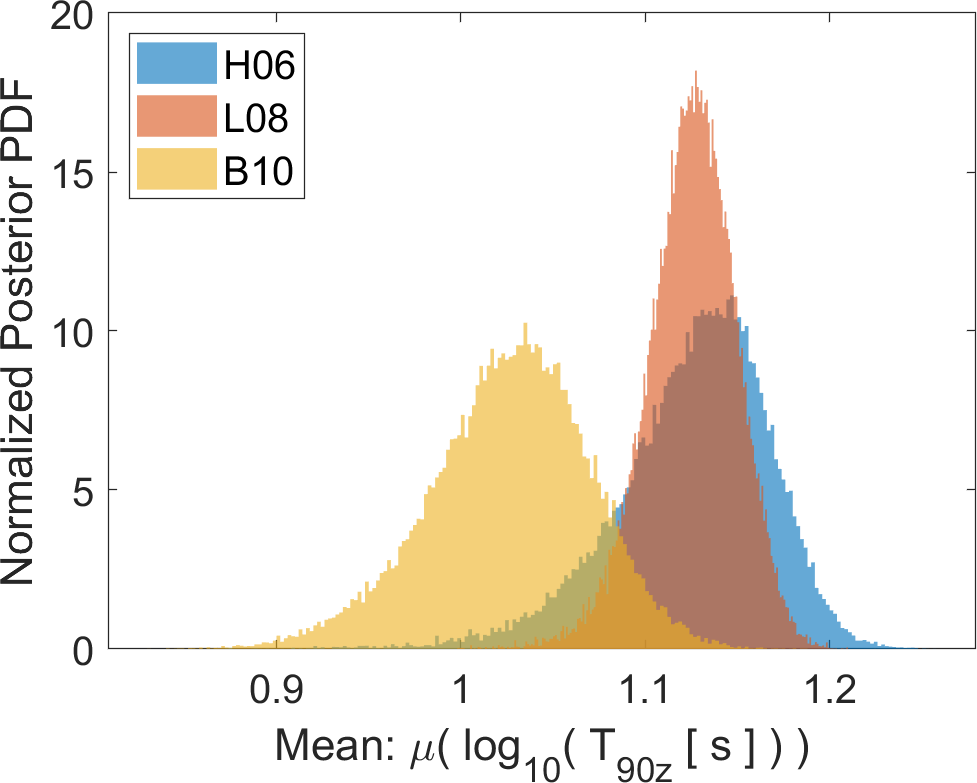} \\
            \includegraphics[width=0.233\textwidth]{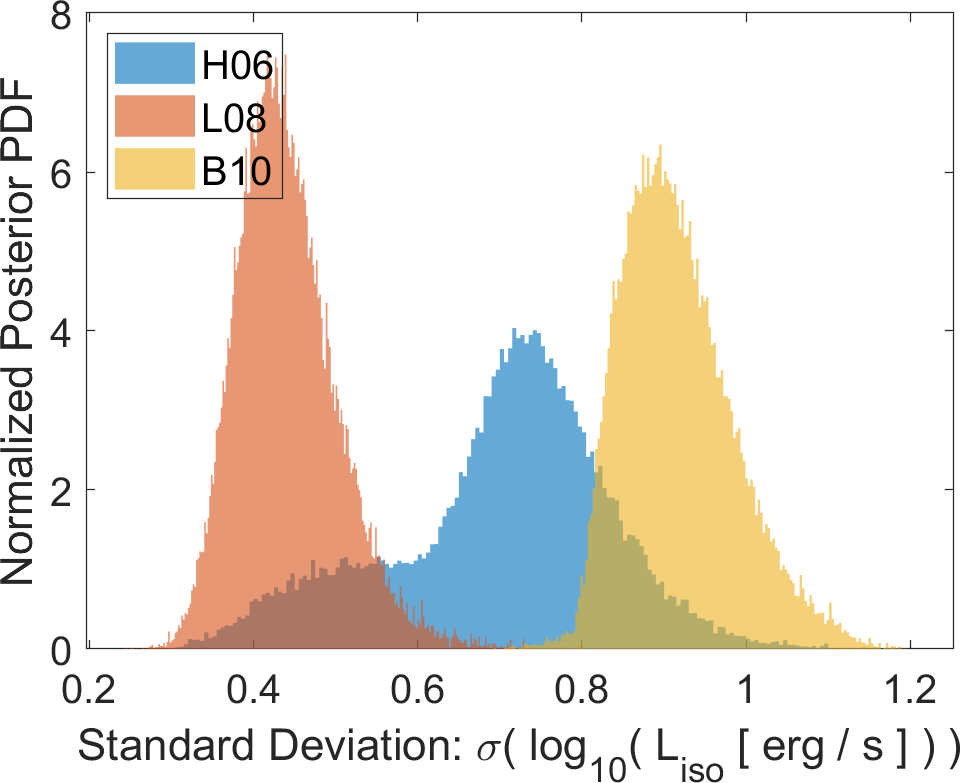} &
            \includegraphics[width=0.233\textwidth]{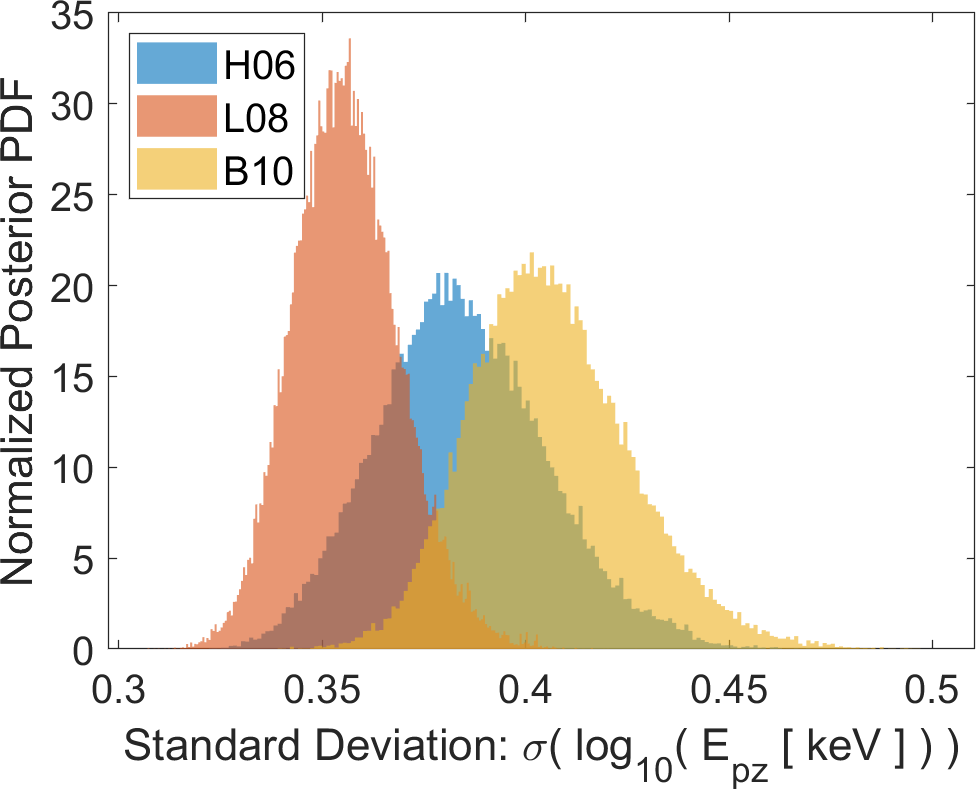} &
            \includegraphics[width=0.233\textwidth]{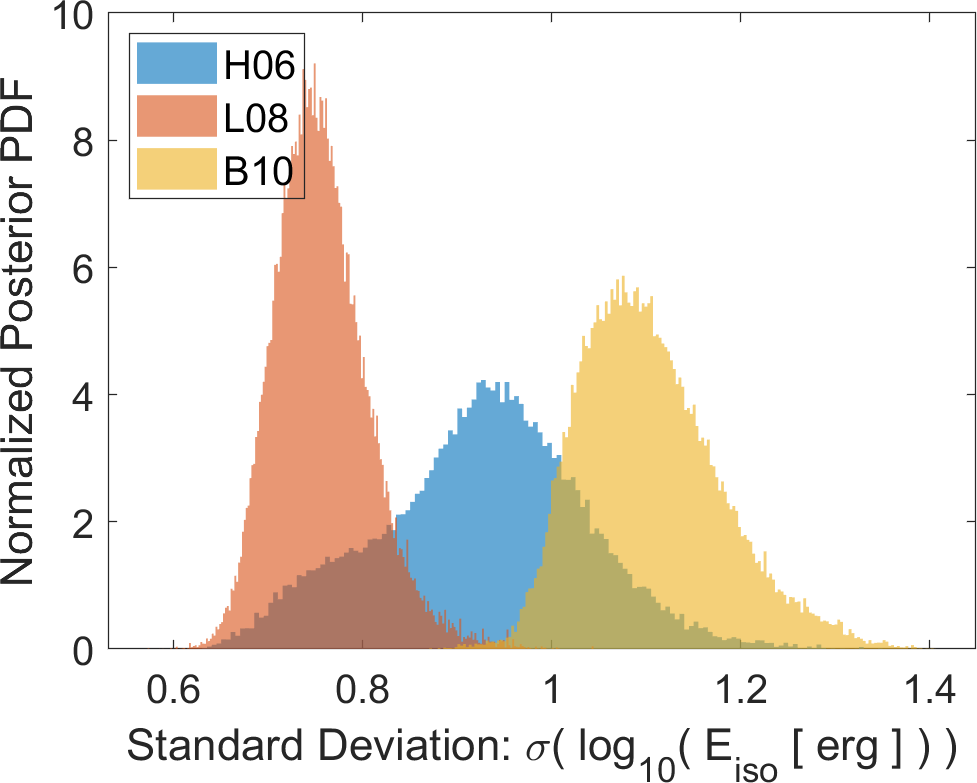} &
            \includegraphics[width=0.233\textwidth]{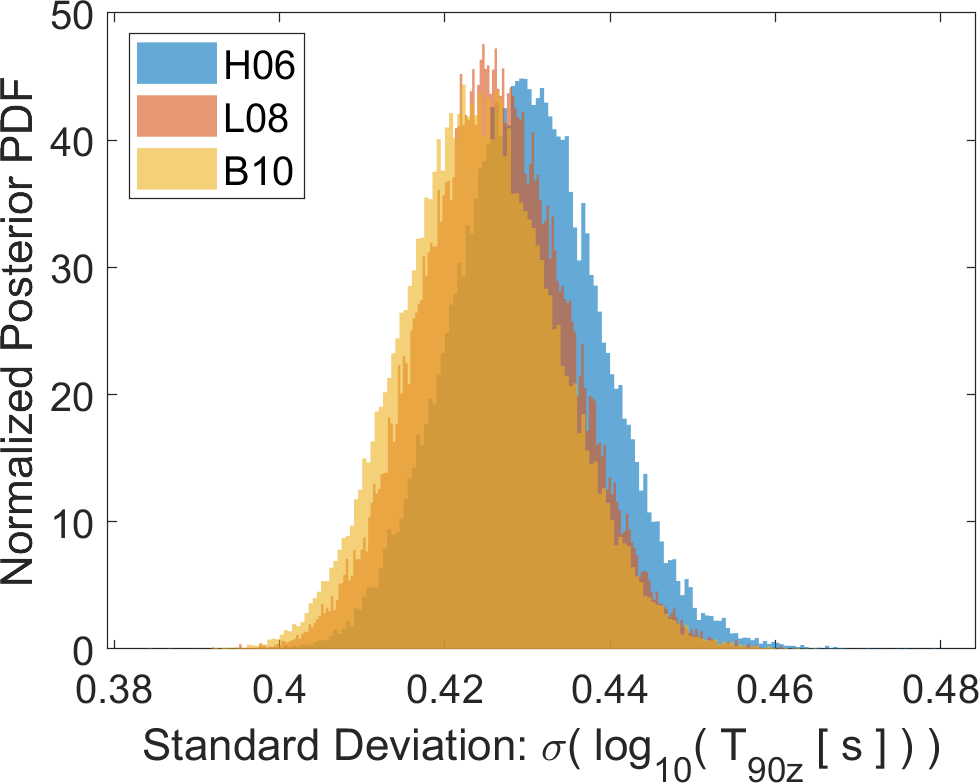} \\
            \includegraphics[width=0.233\textwidth]{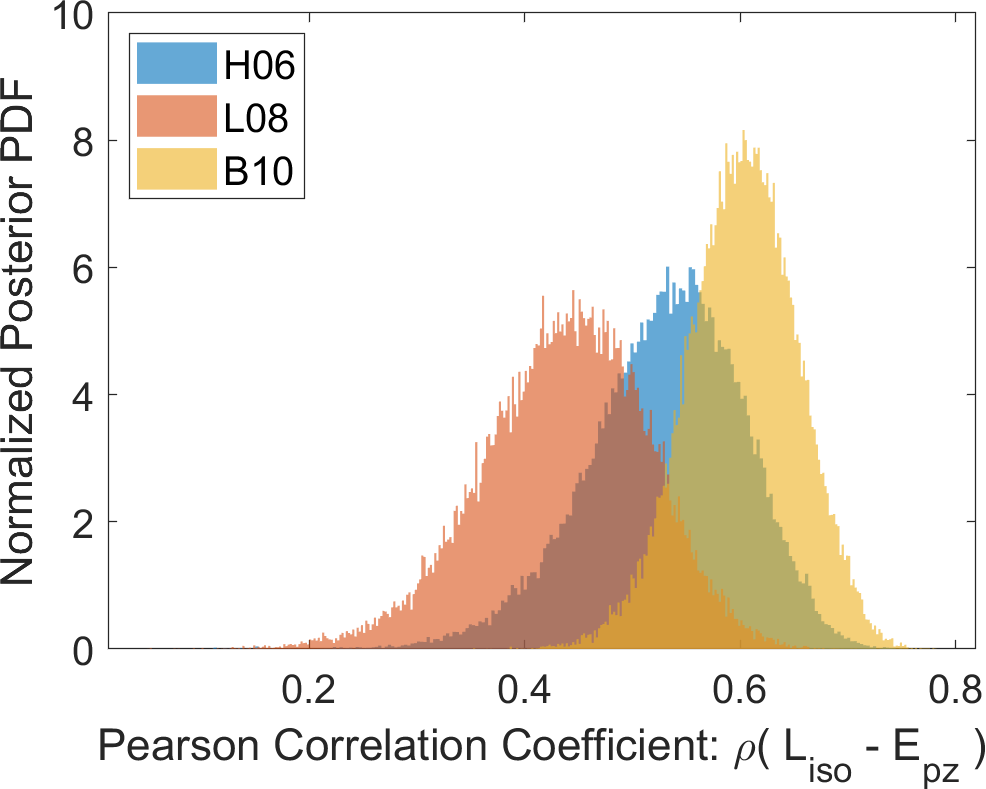} &
            \includegraphics[width=0.233\textwidth]{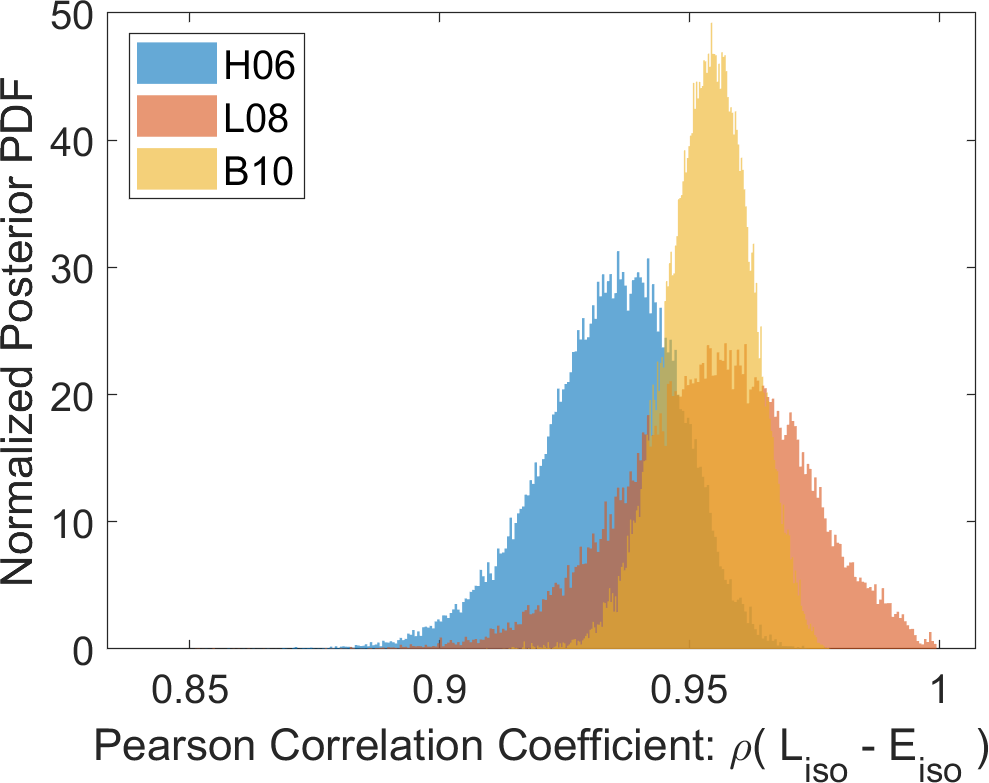} &
            \includegraphics[width=0.233\textwidth]{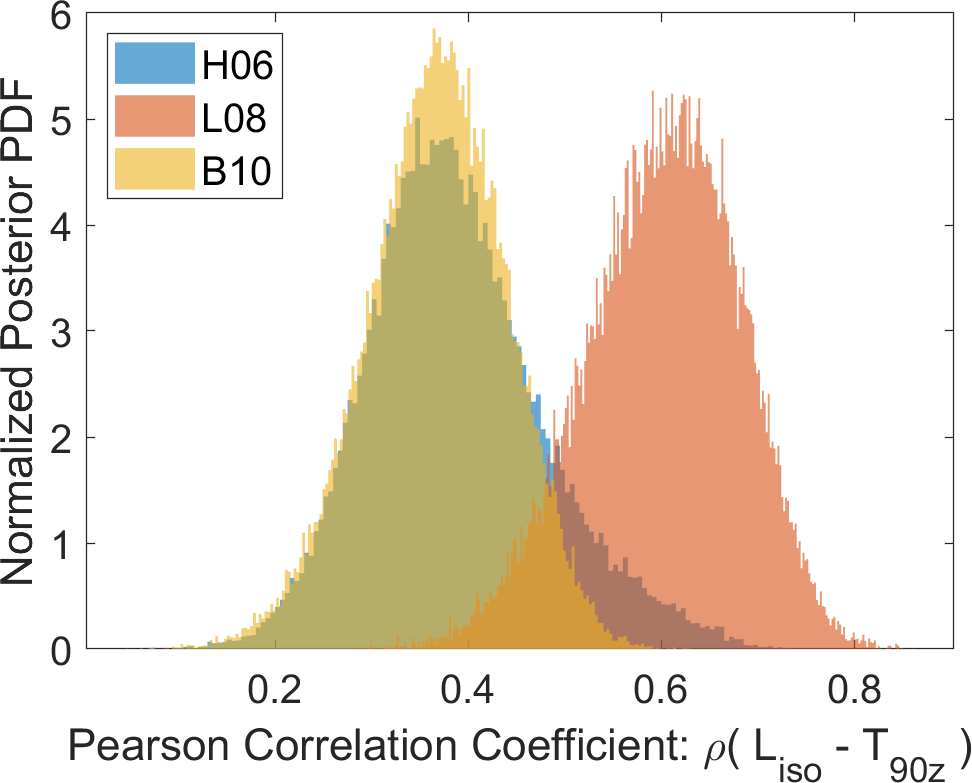} &
            \includegraphics[width=0.233\textwidth]{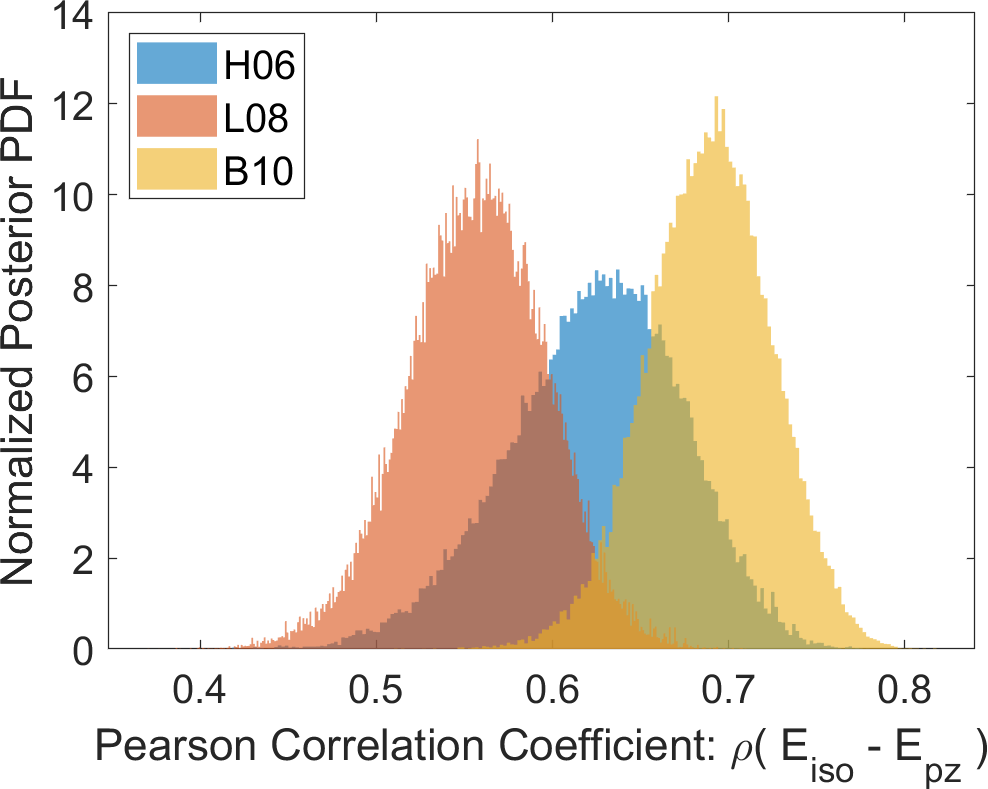} \\
            \includegraphics[width=0.233\textwidth]{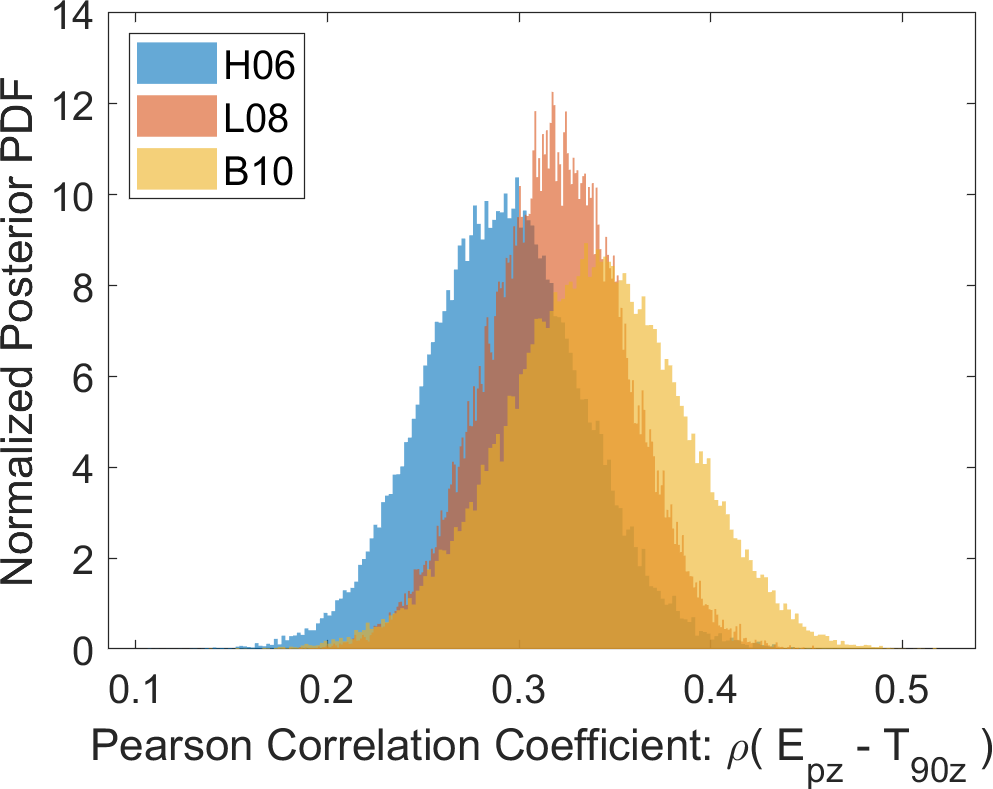} &
            \includegraphics[width=0.233\textwidth]{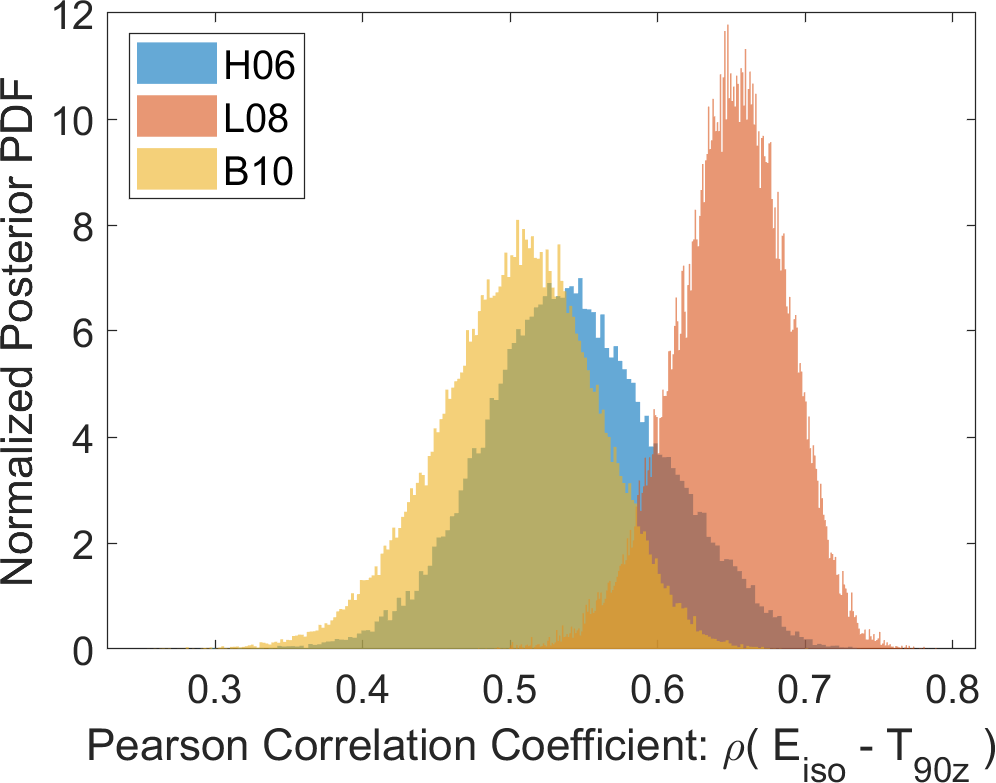} &
            \includegraphics[width=0.233\textwidth]{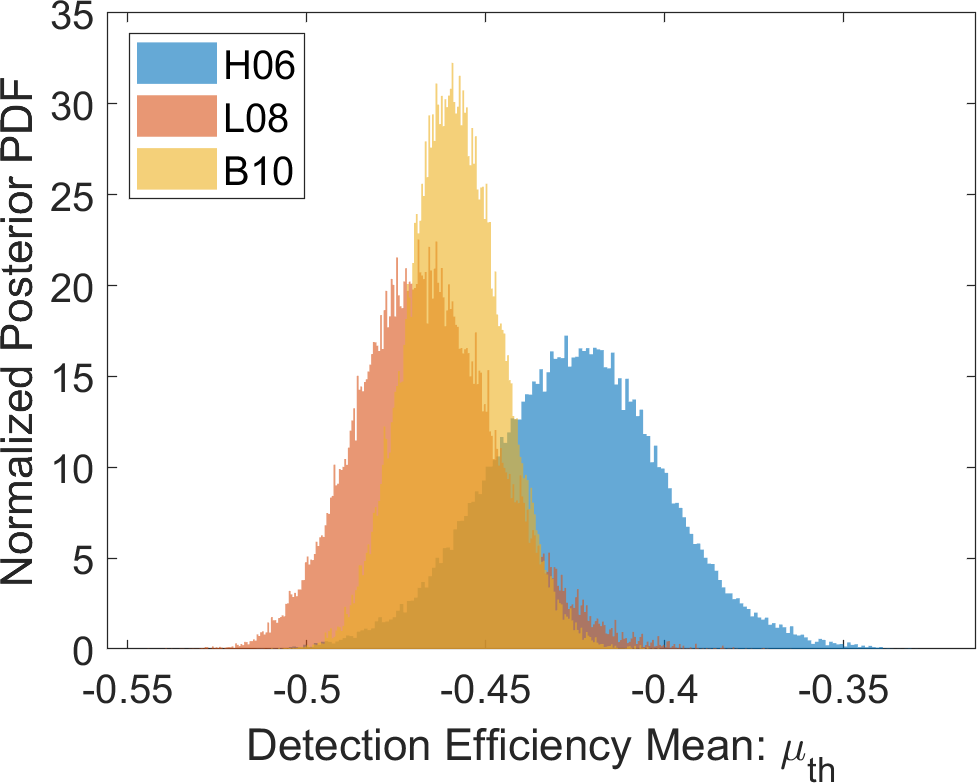} &
            \includegraphics[width=0.233\textwidth]{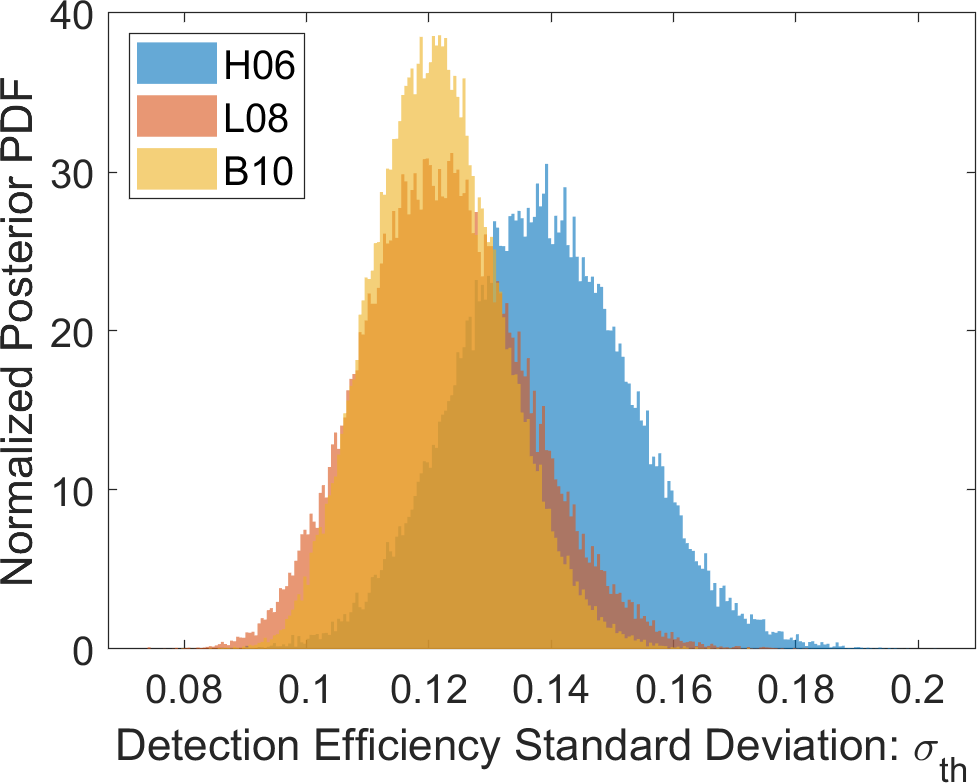} \\
        \end{tabular}
        \caption{The marginal posterior distribution of the 16 parameters of the LGRB world model, for the three redshift distributions considered in this work: \citetalias{hopkins2006normalization}, \citetalias{li2008star}, \citetalias{butler2010cosmic}, as given by \eqref{eq:mz} and \eqref{eq:pz}.\label{fig:paraPostMarginal}}
    \end{figure*}

    \begin{table}[h!]
        \begin{center}
            \vspace{5mm}
            \caption{Mean best-fit parameters of the LGRB World Model, \eqref{eq:modelGeneric}, for the three redshift distribution scenarios considered.\label{tab:paraPostStat}}
            \begin{tabular}{c c c c}
                \hline
                \hline
                Parameter & \citetalias{hopkins2006normalization} & \citetalias{li2008star} & \citetalias{butler2010cosmic} \\
                \hline
                \multicolumn{4}{c}{Redshift Parameters (Equation \ref{eq:mz})} \\
                \hline
                $z_0$      & $0.97$ & $0.993$   & $0.97$  \\
                $z_1$      & $4.5$  & $3.800$   & $4.00$  \\
                $\gamma_0$ & $3.4$  & $3.300$   & $3.14$  \\
                $\gamma_1$ & $-0.3$ & $0.055$   & $1.36$  \\
                $\gamma_2$ & $-7.8$ & $-4.46$   & $-2.92$ \\
                \hline
\multicolumn{4}{c}{Location Parameters($\Mean$)} \\
\hline
$\mu_{\logten(\liso)}$ & $51.07\pm0.33$ & $51.74\pm0.12$ & $51.25\pm0.20$ \\
$\mu_{\logten(\epkz)}$ & $2.36\pm0.09$ & $2.52\pm0.04$ & $2.41\pm0.06$ \\
$\mu_{\logten(\eiso)}$ & $51.54\pm0.34$ & $52.18\pm0.12$ & $51.59\pm0.22$ \\
$\mu_{\logten(\durz)}$ & $1.13\pm0.04$ & $1.13\pm0.02$ & $1.03\pm0.04$ \\
\hline
\multicolumn{4}{c}{Scale Parameters (diagonal elements of $\CovMat$)} \\
\hline
$\sigma_{\logten(\liso)}$ & $0.70\pm0.13$ & $0.44\pm0.06$ & $0.92\pm0.07$ \\
$\sigma_{\logten(\epkz)}$ & $0.38\pm0.02$ & $0.36\pm0.01$ & $0.41\pm0.02$ \\
$\sigma_{\logten(\eiso)}$ & $0.93\pm0.11$ & $0.76\pm0.05$ & $1.10\pm0.07$ \\
$\sigma_{\logten(\durz)}$ & $0.43\pm0.01$ & $0.43\pm0.01$ & $0.42\pm0.01$ \\
\hline
\multicolumn{4}{c}{Correlation Coefficients (non-diagonal elements of $\CovMat$)} \\
\hline
$\rho_{\liso-\epkz}$ & $0.53\pm0.07$ & $0.44\pm0.08$ & $0.60\pm0.05$ \\
$\rho_{\liso-\eiso}$ & $0.93\pm0.01$ & $0.95\pm0.02$ & $0.95\pm0.01$ \\
$\rho_{\liso-\durz}$ & $0.39\pm0.09$ & $0.60\pm0.08$ & $0.37\pm0.07$ \\
$\rho_{\epkz-\eiso}$ & $0.62\pm0.05$ & $0.56\pm0.04$ & $0.69\pm0.03$ \\
$\rho_{\epkz-\durz}$ & $0.29\pm0.04$ & $0.32\pm0.04$ & $0.34\pm0.05$ \\
$\rho_{\eiso-\durz}$ & $0.54\pm0.06$ & $0.65\pm0.04$ & $0.50\pm0.05$ \\
\hline
\multicolumn{4}{c}{BATSE Detection Efficiency (Sample Incompleteness)} \\
\hline
$\mu_{th}$ & $-0.42\pm0.02$ & $-0.47\pm0.02$ & $-0.46\pm0.01$ \\
$\mu_{th}$ & $0.14\pm0.01$ & $0.12\pm0.01$ & $0.12\pm0.01$ \\
                \hline
                \hline
            \end{tabular}
        \end{center}
        {Note.--- The joint posterior distribution of the 16-dimensional parameters of the model resulting from the Markov chains are available for download at \url{https://github.com/shahmoradi/BatseRedshiftEstimates} for each of the three redshift distributions.}
    \end{table}

    We proceed by first fitting the proposed model to 1366 BATSE LGRB data under the three LGRB redshift distribution scenarios prescribed by \eqref{eq:mz} and \eqref{eq:pz}. For each LGRB rate density scenario, the posterior PDF of parameters in \eqref{eq:paraPostPoisson} is explored by an adaptive Metropolis-Hastings Markov Chain Monte Carlo algorithm which we have developed for such sampling tasks \citep[e.g.,][]{shahmoradi2013multivariate, shahmoradi2013gamma, shahmoradi2014similarities, shahmoradi2015short, shahmoradi2017lgrbworldmodel, shahmoradi2017sgrbworldmodel, shahmoradi2018multilevel, shahmoradi2019paramonte}. The best-fit parameters corresponding to the three SFR densities are tabulated in Table \ref{tab:paraPostStat}, and their marginal distributions for the three LGRB rate densities are compared in Figure \ref{fig:paraPostMarginal}.\newpar

    \begin{figure*}[t!]
        \centering
        \begin{tabular}{ccc}
            \includegraphics[width=0.316\textwidth]{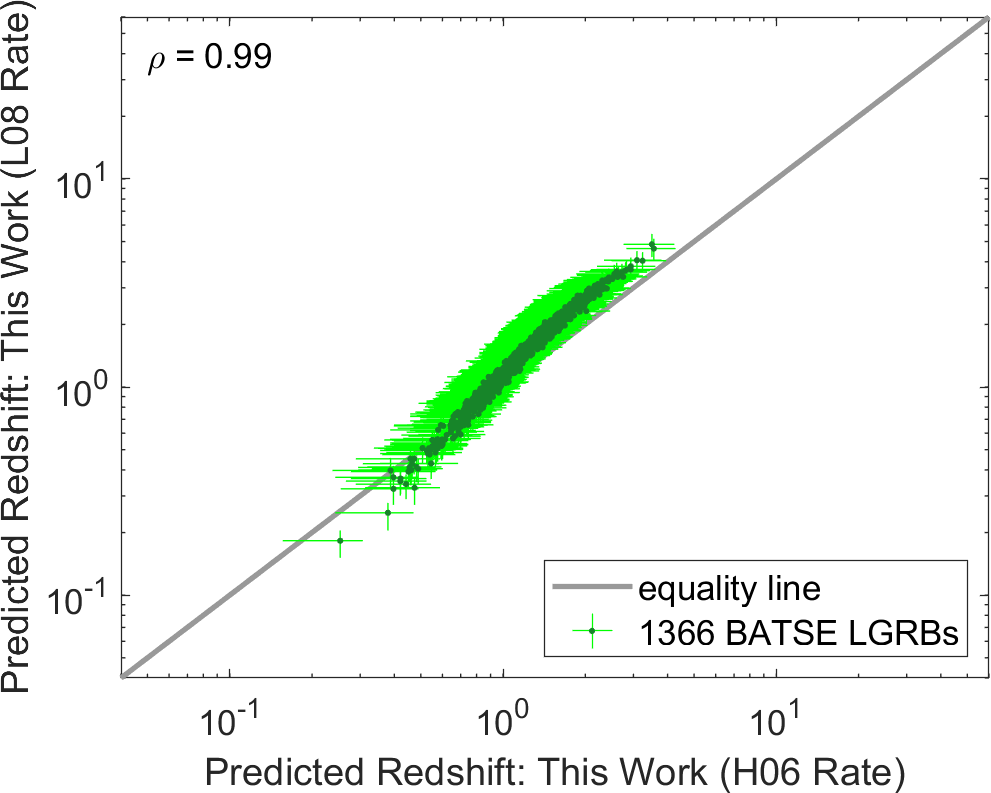} &
            \includegraphics[width=0.316\textwidth]{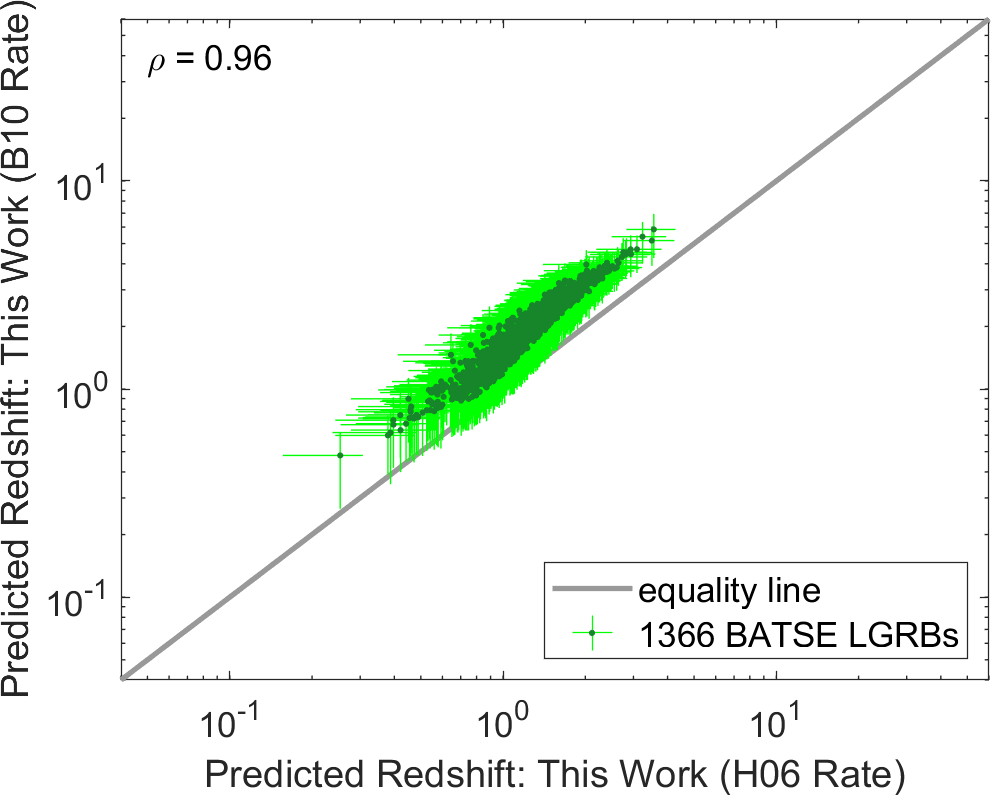} &
            \includegraphics[width=0.316\textwidth]{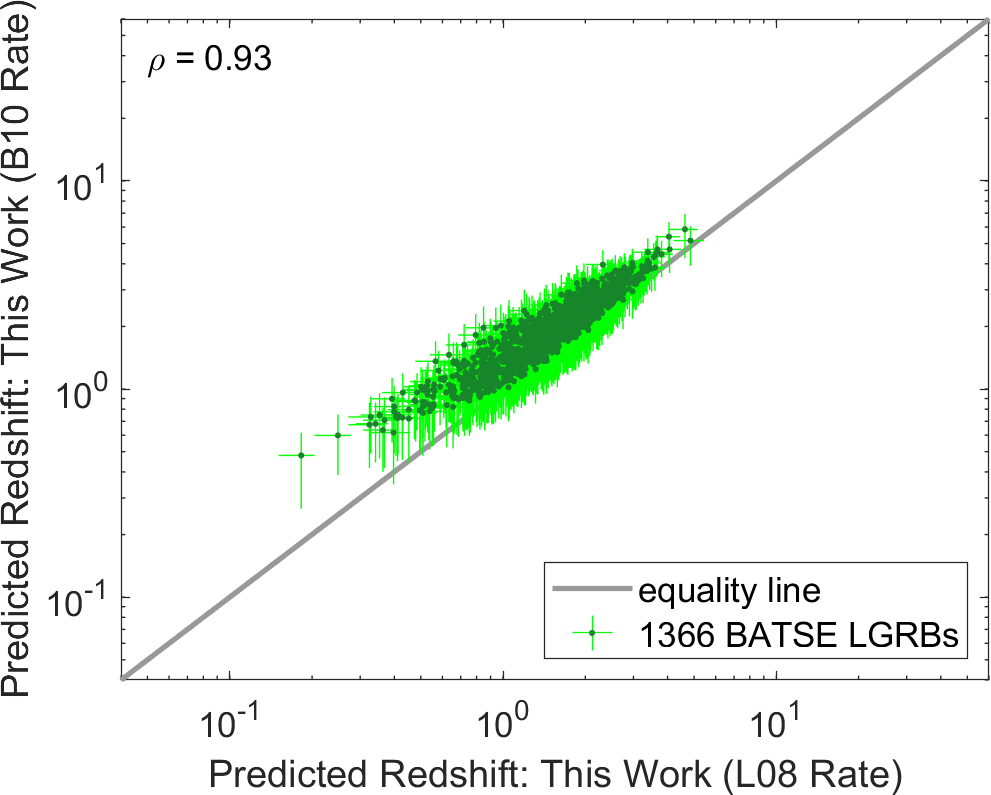} \\
        \end{tabular}
        \caption{A comparison of the expected redshifts of 1366 BATSE LGRBs for the three different cosmic LGRB rate density assumptions \eqref{eq:mz} with each other. On each plot, the Pearson's correlation coefficient of the two sets of expected redshifts is also reported. The error bars represent the $50\%$ prediction intervals of each individual redshift.
        \label{fig:expectedRedshift}}
    \end{figure*}

    Once the parameters of the model are constrained, we use the calibrated model, \eqref{eq:modelobs}, at the second level of our analysis to further constrain the PDFs of the unknown redshifts of individual BATSE LGRBs according to \eqref{eq:zMarginalPDF}. Similar to the Empirical Bayes methodology, this iterative process can continue until convergence to a specific set of redshift PDFs occurs. Given the computational complexity and expense of each iteration however, we stop after the first round of estimates, which is also a common practice in the Empirical Bayes modeling. In addition, we further reduce the computational complexities of the inverse problems by dropping the uncertainties in all observational data from both levels of the analysis in \eqref{eq:paraPostPoisson} and \eqref{eq:zMarginalPDF}. To further reduce the computational expense of the inference, we also approximate the numerical integration in the definition of the luminosity distance in \eqref{eq:ldis} by the analytical expressions of \citet{wickramasinghe2010analytical}. All routines were implemented in Fortran programming language and comply with the latest Fortran Standard in 2018 \citep[e.g.,][]{metcalf2011modern, metcalf2018modern}.\newpar

    The resulting expected redshifts of 1366 BATSE LGRBs for the rate density assumptions of \citetalias{hopkins2006normalization}, \citetalias{li2008star}, and \citetalias{butler2010cosmic} are compared with each other in Figure \ref{fig:expectedRedshift}. The mean redshifts together with $50\%$ and $90\%$ prediction intervals for the three rate density scenarios are also reported in Table \ref{tab:redshiftCatalog}. For the sake of completeness, we also compile and report in Table \ref{tab:BatseData}, the set of BATSE LGRB spectral and temporal properties used for the calibration of the models.\newpar

    On average, the redshifts of individual BATSE LGRBs can be constrained to within 50\% uncertainty ranges of 0.60432, 0.37012, and 1.075 corresponding to the three LGRB rate densities of \citetalias{hopkins2006normalization}, \citetalias{li2008star}, and \citetalias{butler2010cosmic} respectively. At 90\% confidence level, the prediction intervals expand to wider uncertainty ranges of 1.5466, 0.97878, and 2.6498 respectively.

\section{discussion}
\label{sec:discussion}

    \begin{figure}[t!]
        \centering
        \includegraphics[width=0.45\textwidth]{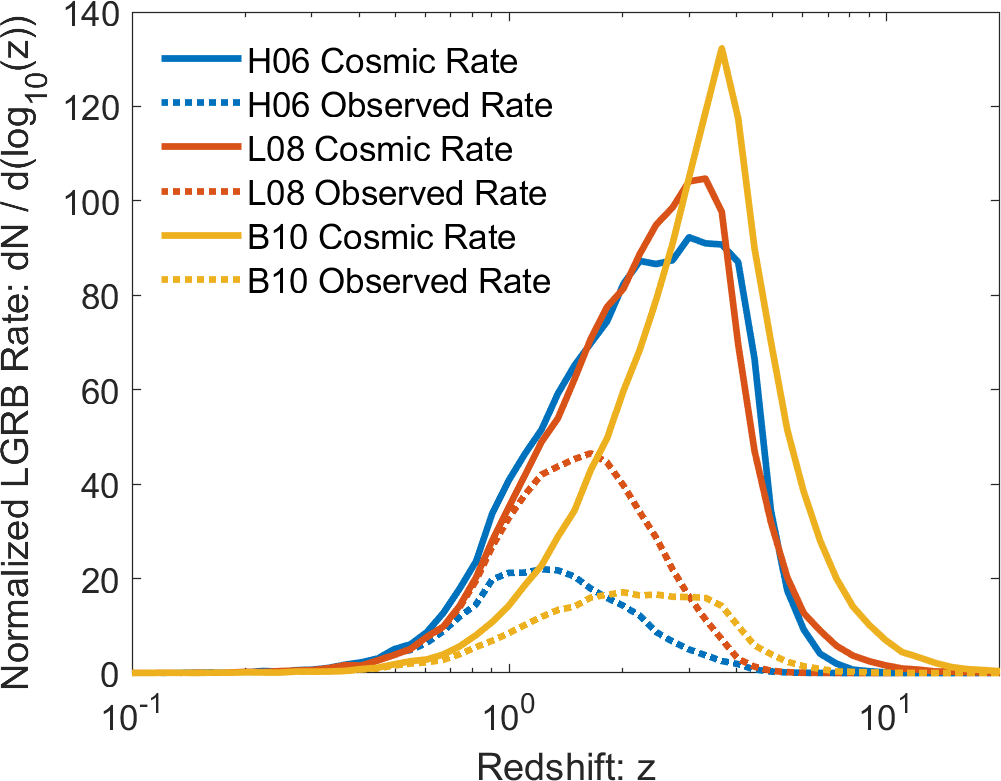}
        \caption{A comparison of the prescribed cosmic LGRB rates (represented by the solid lines) according to the three models of \citetalias{hopkins2006normalization}, \citetalias{li2008star}, and \citetalias{butler2010cosmic} as given by \eqref{eq:mz} \& \eqref{eq:pz} and their corresponding observed rates (represented by the dashed lines) as detected by BATSE LADs. Note that the rates are computed in $\logten(z+1)$ space.\label{fig:histRedshift}}
    \end{figure}

    In this work, we proposed a semi-Bayesian methodology to infer the unknown redshifts of 1366 BATSE catalog LGRBs. Towards this, first, we segregated the two populations of BATSE LGRBs and SGRBs using a fuzzy classification based on the observed duration and spectral peak energies of 1966 BATSE GRBs with available spectral and temporal information. We then modeled the process of LGRB detection as a non-homogeneous spatiotemporal Poisson process, whose rate parameter was modeled by a multivariate log-normal distribution as a function of the four main LGRB intrinsic attributes: the 1024 [ms] isotropic peak luminosity ($\liso$), the total isotropic emission ($\eiso$), the intrinsic spectral peak energy ($\epkz$), and the intrinsic duration ($\durz$). In order to calibrate the parameters of the rate model, we made a fundamental assumption that LGRBs trace the Cosmic Star Formation Rate (SFR), or a metallicity-corrected SFR. For each of the three individual LGRB rate densities assumed in this work: \citetalias{hopkins2006normalization}, \citetalias{li2008star}, \citetalias{butler2010cosmic}, we then used the resulting posterior probability densities of the model parameters to compute the probability density functions of the redshifts of individual BATSE LGRBs.\newpar

    As illustrated in Figure \ref{fig:expectedRedshift}, we find that the individual redshift estimates of BATSE LGRBs for the LGRB rate density assumptions are broadly consistent with each other. There are, however, systematic differences between the three estimates, the most significant of which is the difference between the expected redshifts based on \citetalias{butler2010cosmic} and the two other rate densities. The difference can be explained by the larger rates of LGRBs that the model of \citetalias{butler2010cosmic} implies at higher redshifts compared to \citetalias{hopkins2006normalization} and \citetalias{li2008star}. This, in effect, shifts the expected redshifts of BATSE LGRBs systematically toward larger values. The excess of LGRB occurrence rate at high redshifts in the model of \citetalias{butler2010cosmic} is also well illustrated in Figure \ref{fig:histRedshift}.\newpar

    The underlying logic of our approach to resolving the individual redshifts of BATSE LGRBs can be qualitatively understood by reconsidering the set of equations in \eqref{eq:obsIntMap}. Taking the logarithm of both sides of all equations, we obtain a set of linear maps from the rest-frame to the observer-frame properties of LGRBs. Therefore, the distributions of the four observer-frame LGRB properties result from the convolution of the distributions of the corresponding rest-frame LGRB properties with the distribution of terms that are exactly determined by redshift (i.e., the logarithm of the luminosity distance, $\logten(\ldis)$ and the term $\logten(z+1)$).\newpar

    \begin{figure*}[t!]
        \centering
        \begin{tabular}{ccc}
            \includegraphics[width=0.316\textwidth]{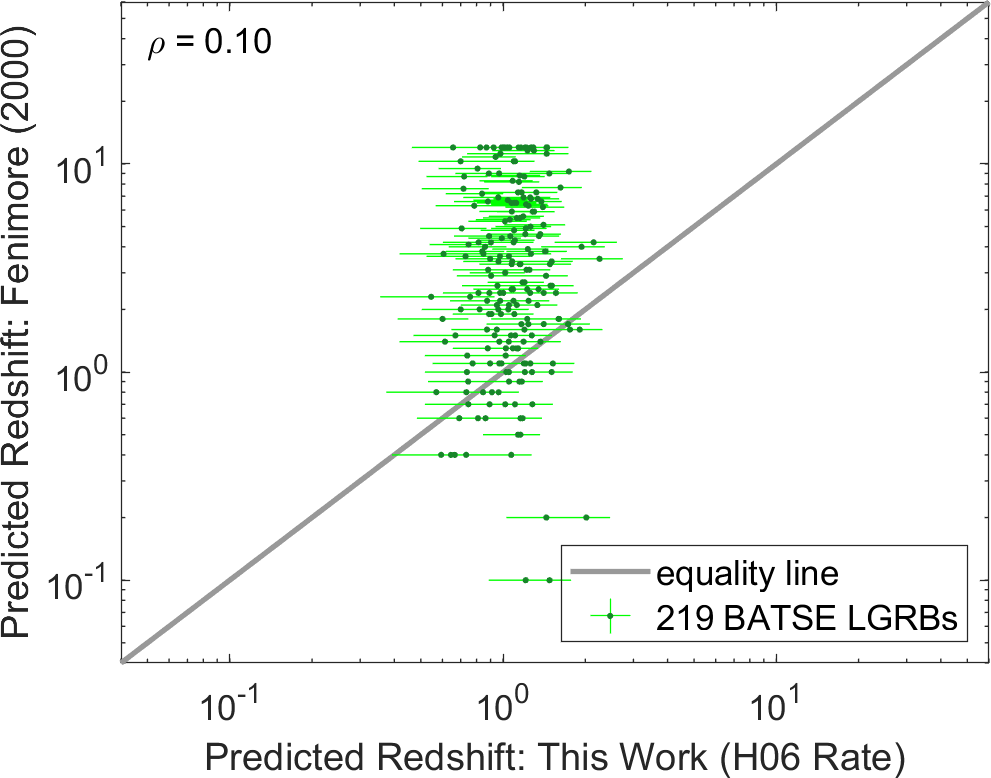} & \includegraphics[width=0.316\textwidth]{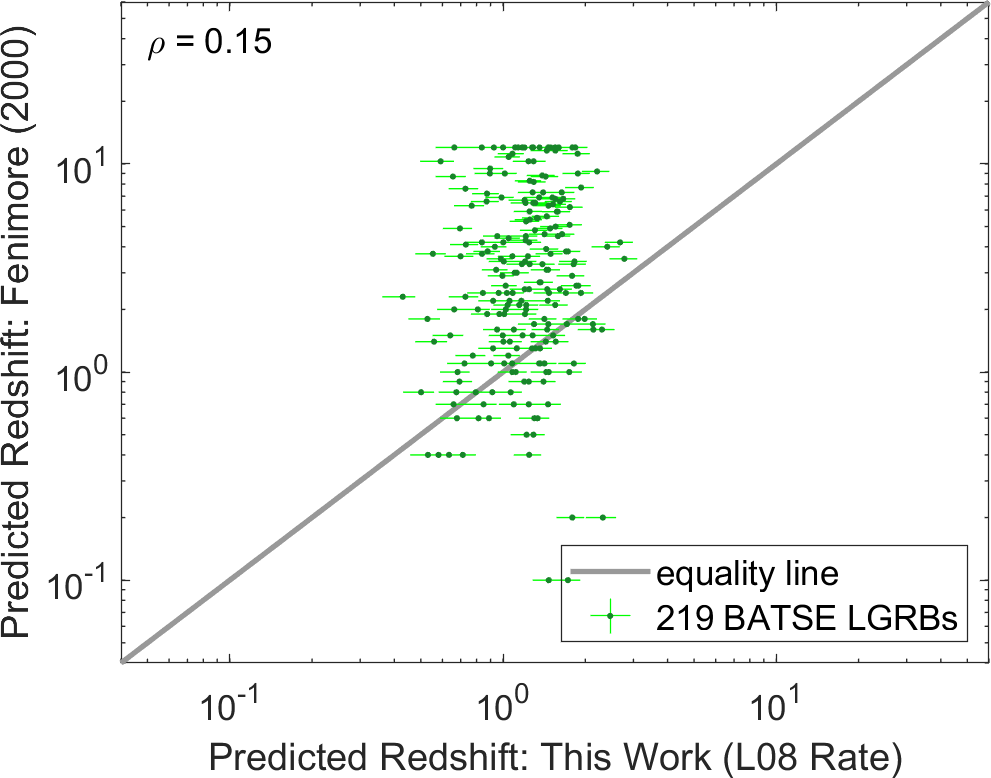} & \includegraphics[width=0.316\textwidth]{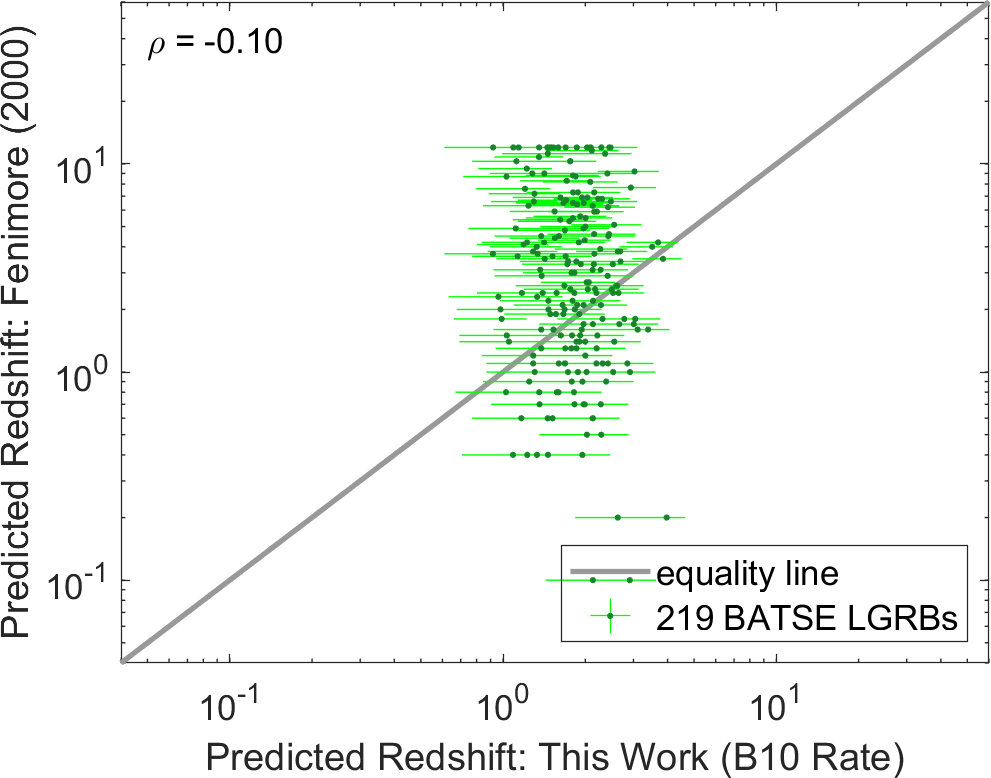} \\
            \includegraphics[width=0.316\textwidth]{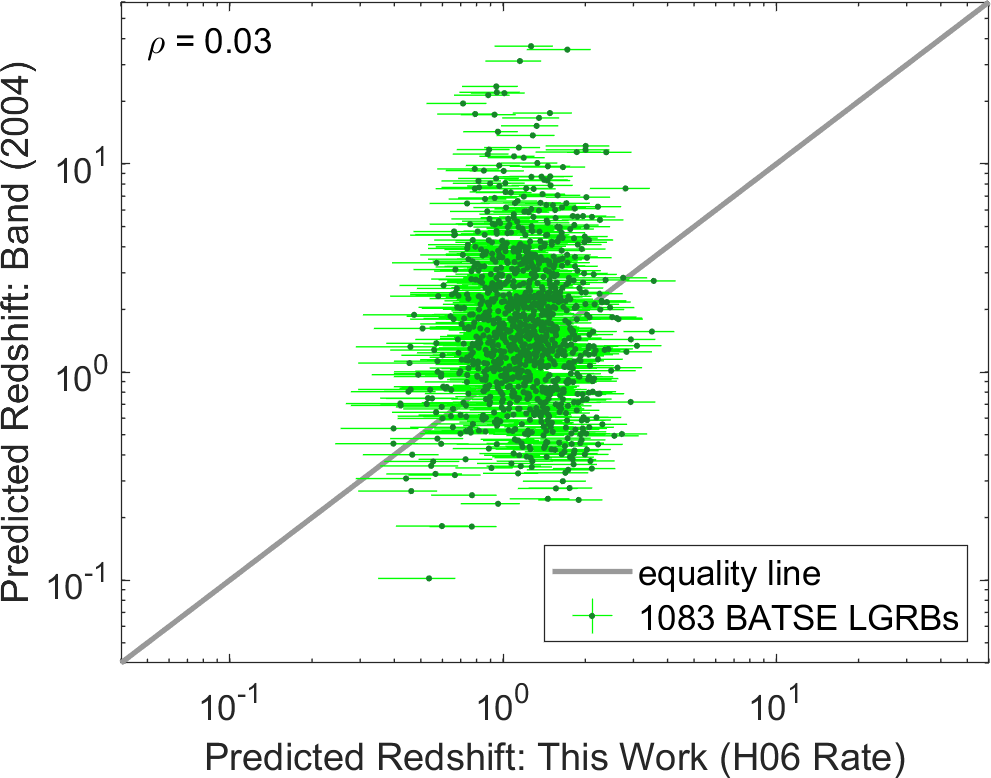} & \includegraphics[width=0.316\textwidth]{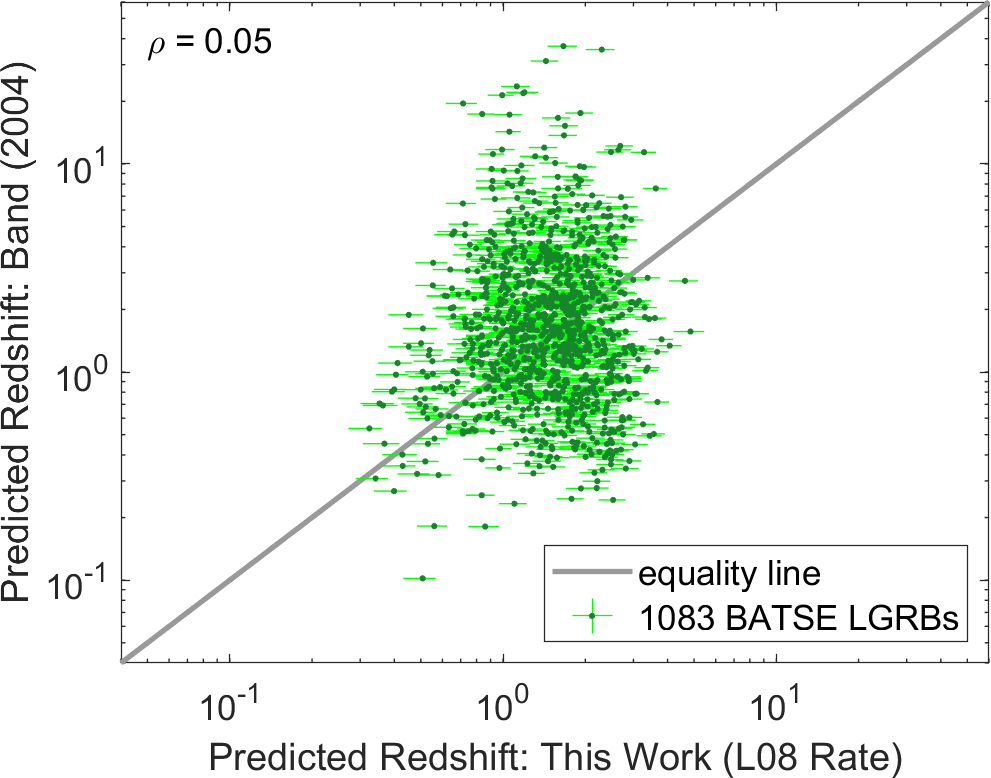} & \includegraphics[width=0.316\textwidth]{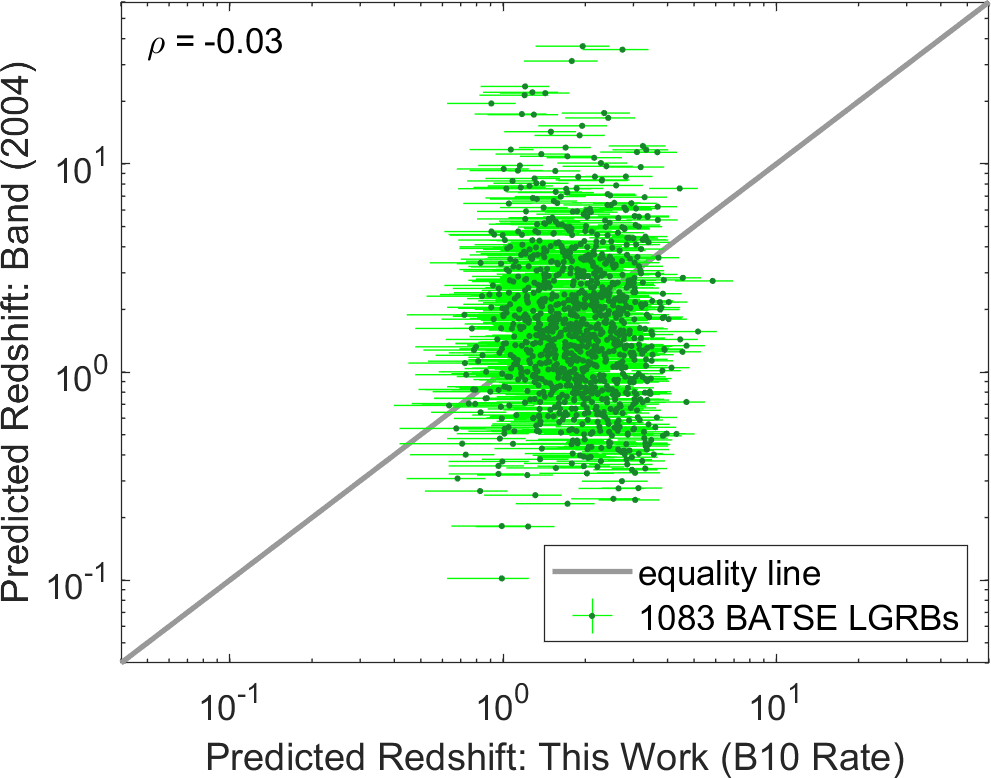} \\
            \includegraphics[width=0.316\textwidth]{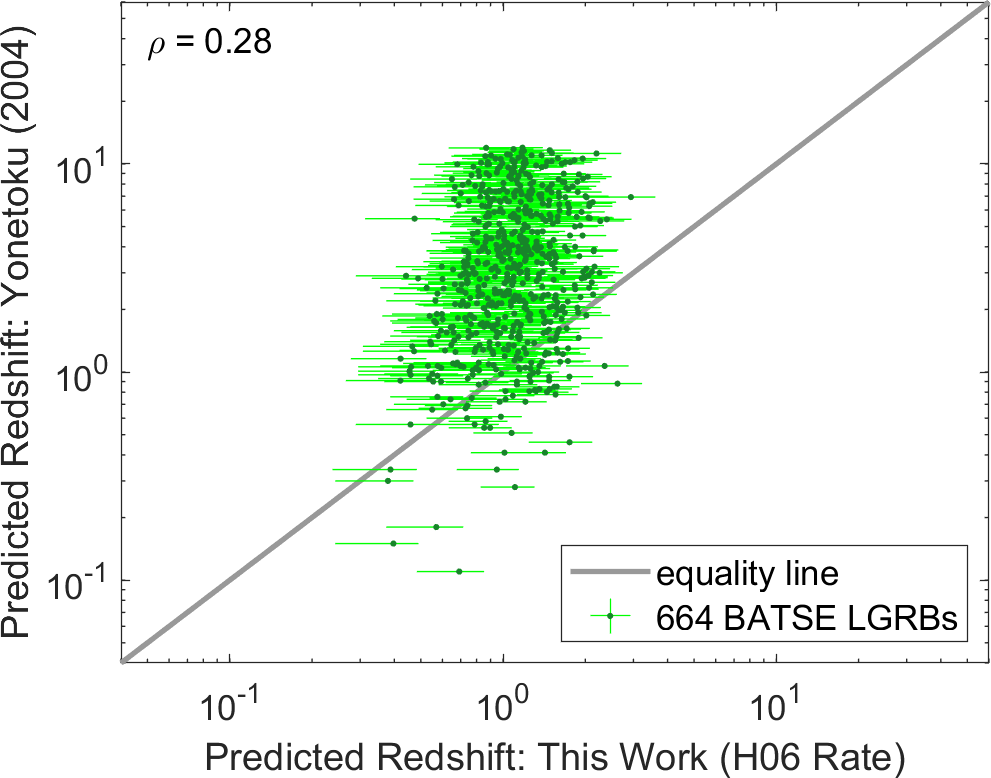} & \includegraphics[width=0.316\textwidth]{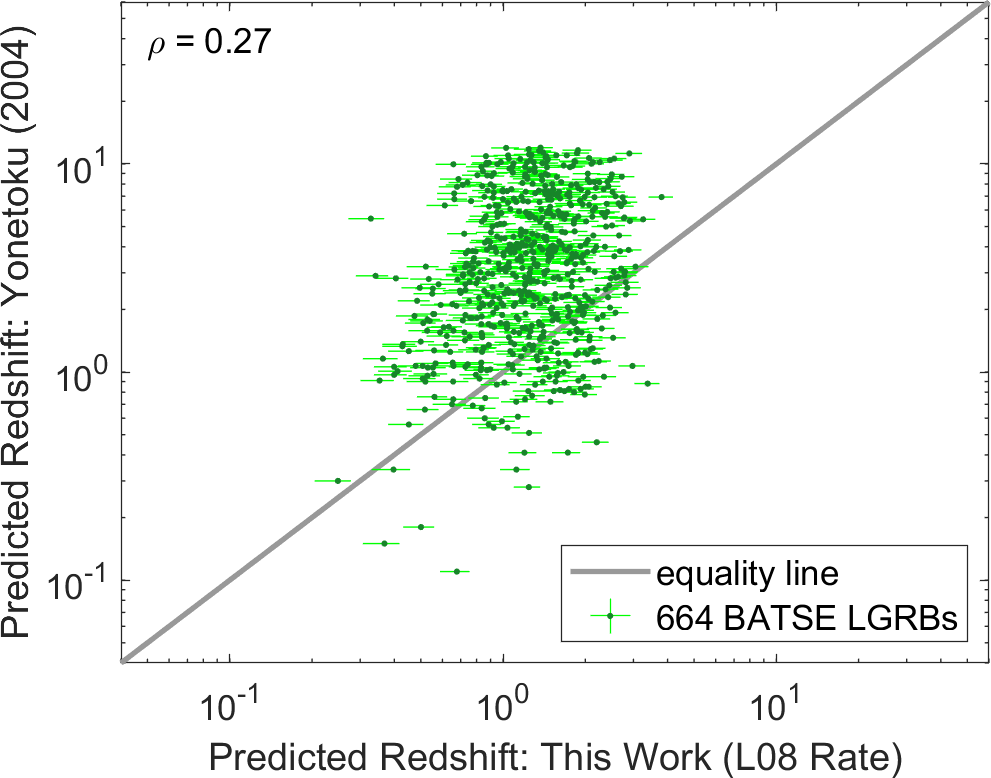} & \includegraphics[width=0.316\textwidth]{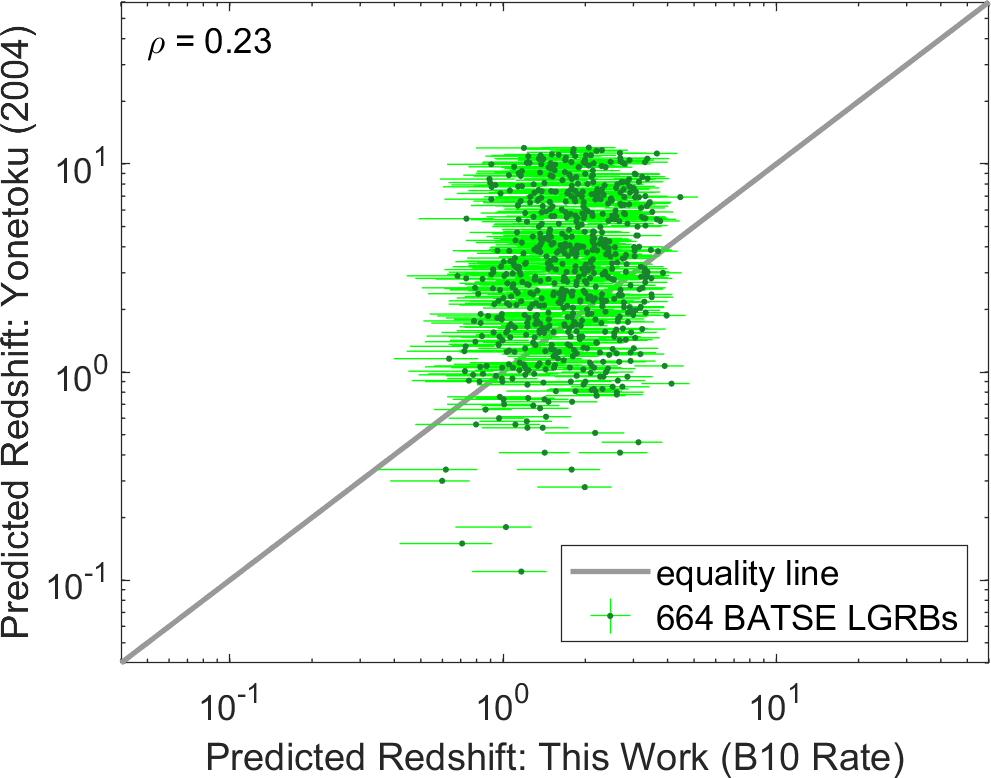} \\
        \end{tabular}
        \caption{
        A comparison of the expected redshifts of BATSE LGRBs under the three different cosmic LGRB rate densities assumptions \eqref{eq:mz} with the BATSE redshift estimates from previous works, based on the correlation between lightcurve-variability and the peak luminosity by \citet{fenimore2000redshifts}, the lag-luminosity relation by \citet{band2004gamma}, and the correlation between the peak luminosity and the spectral peak energy by \citet{yonetoku2004gamma}. On each plot, the Pearson's correlation coefficient of the two sets of expected redshifts is also reported. The error bars represent the $50\%$ predictions intervals for each predicted redshift in this work.\label{fig:redshiftComparisonWithOldWorks}}
    \end{figure*}

    The larger the variances of redshift-related terms in these equations are (compared to the variances of rest-frame LGRB attributes), the more the observer-frame LGRB properties will be indicative of the redshifts of individual events. For example, the LGRB rate density of \citetalias{li2008star} results in the largest ratios of the variances of redshift-related terms \eqref{eq:obsIntMap} to the variances of the intrinsic BATSE LGRB attributes. This, in turn, leads to the least uncertain (but not necessarily the most accurate) individual redshift predictions under the rate density assumptions of \citetalias{li2008star} among the three rate densities considered in this study. This is also evidenced in the plots of Figure \ref{fig:expectedRedshift} and \ref{fig:redshiftComparisonWithOldWorks} by the relatively tighter prediction intervals (i.e., error bars) for the redshift estimates based on the LGRB rate density of \citetalias{li2008star}.\newpar

    \begin{figure*}[tphb]
        \centering
        \begin{tabular}{ccc}
            \includegraphics[width=0.316\textwidth]{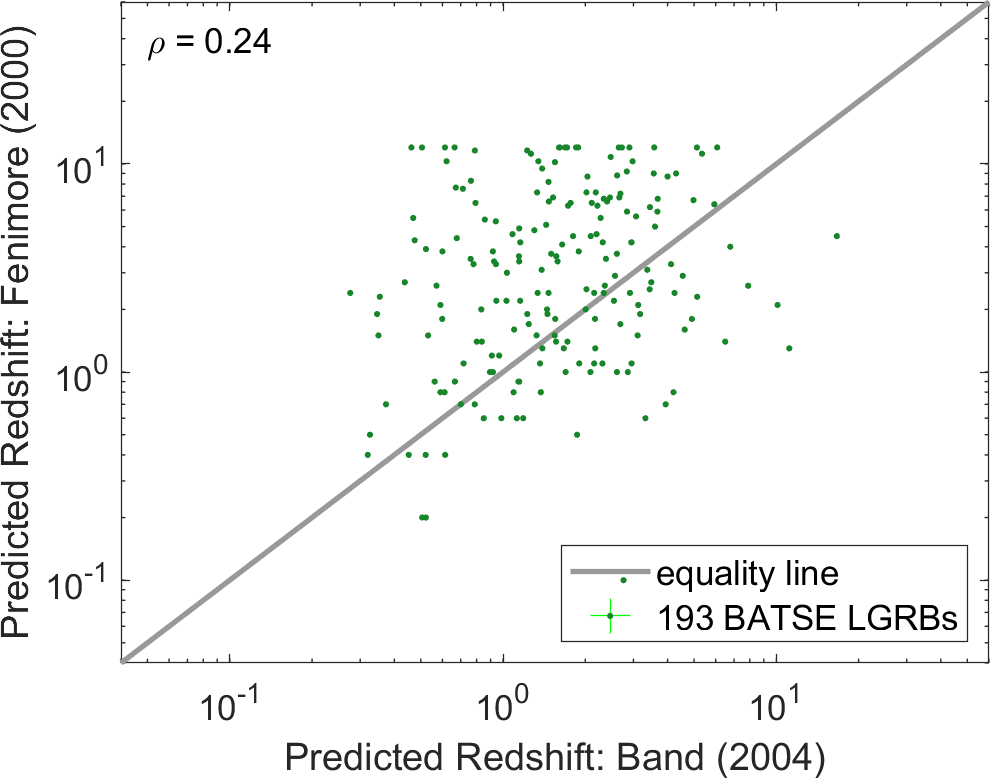} &
            \includegraphics[width=0.316\textwidth]{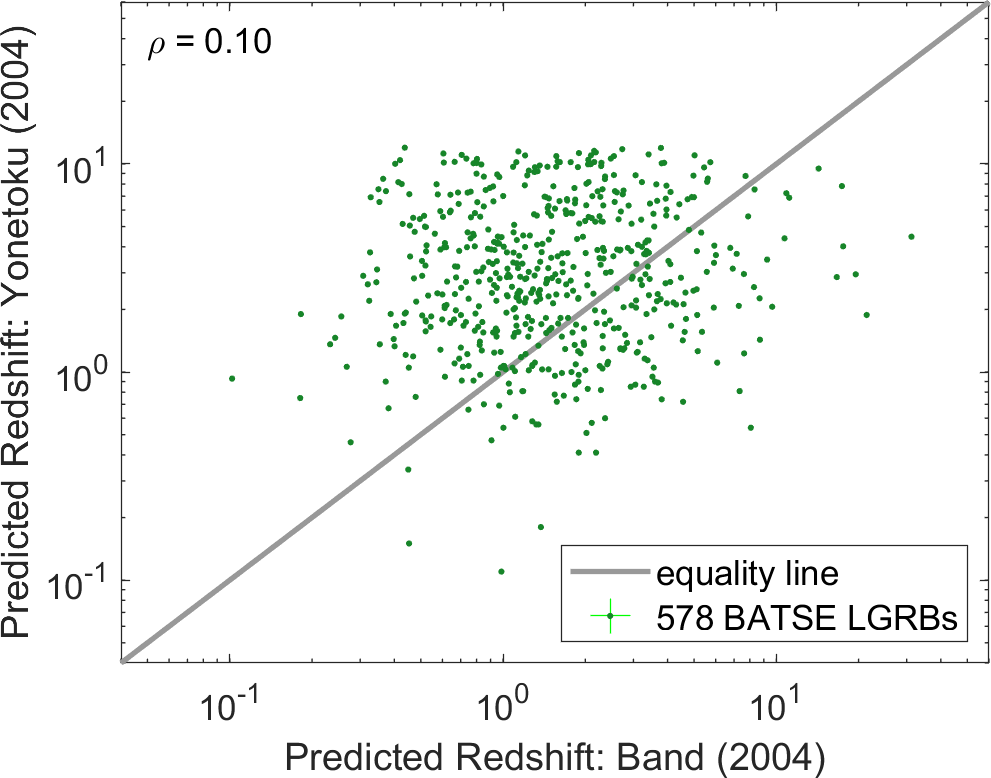} &
            \includegraphics[width=0.316\textwidth]{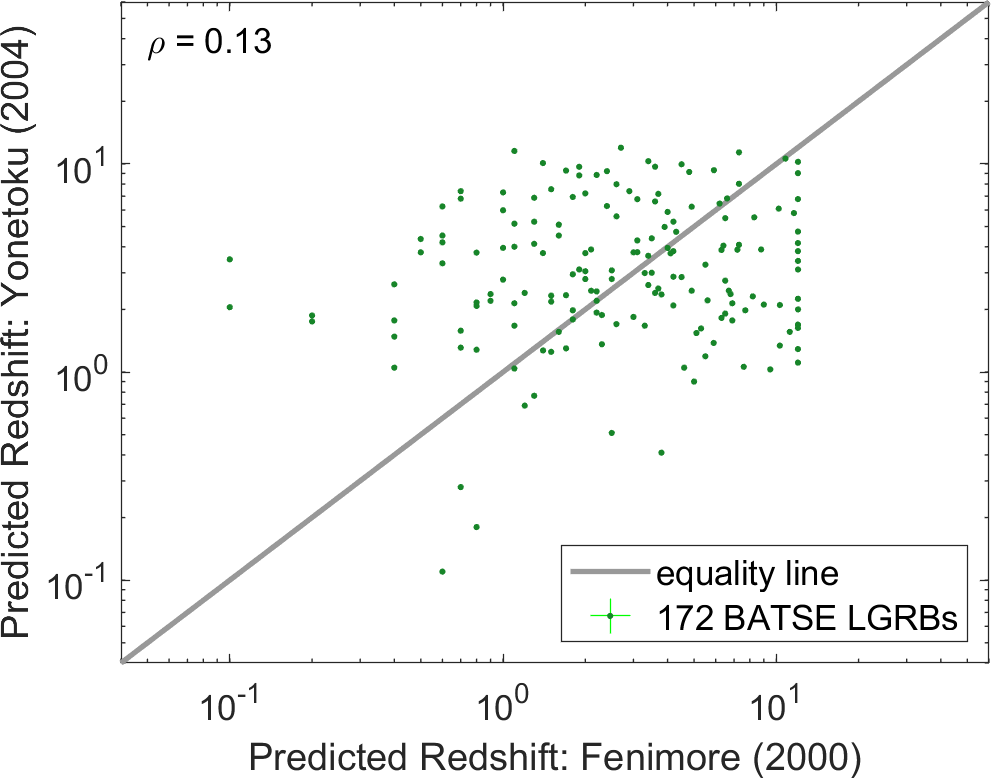} \\
        \end{tabular}
        \caption{A comparison of the predicted redshifts of BATSE GRBs based on the correlation between lightcurve-variability and the peak luminosity by \citet{fenimore2000redshifts}, the lag-luminosity relation by \citet{band2004gamma}, and the correlation between the peak luminosity and the spectral peak energy by \citet{yonetoku2004gamma} with each other. On each plot, the Pearson's correlation coefficient of the two sets of predicted redshifts is also reported.\label{fig:oldWorkComparison}}
    \end{figure*}

    A number of previous studies have attempted to estimate the unknown redshifts of BATSE LGRBs via some of the phenomenological gamma-ray correlations. For example, \citet{fenimore2000redshifts} used the apparent correlation between the isotropic peak luminosity of GRBs and the temporal variability of their lightcurves to estimate redshifts of 220 BATSE GRBs. \citet{band2004gamma} used the apparent correlation between the spectral lag and the peak luminosity of GRBs to estimate the unknown redshifts of 1194 BATSE events. Similarly, \citet{yonetoku2004gamma} used the apparent observed correlation between the isotropic peak luminosity and the intrinsic spectral peak energies of GRBs to estimate redshifts of 689 BATSE GRBs.\newpar

    We compare our redshift estimates for the three LGRB rate densities to the predictions of each of the aforementioned works in Figure \ref{fig:redshiftComparisonWithOldWorks}. None of these three previous independent redshift estimates based on the high-energy correlations appear to agree with our predictions. The three independent estimates are also highly inconsistent with each other as shown in Figure \ref{fig:oldWorkComparison}.\newpar

    The observed inconsistencies of the previous independent redshift estimates of BATSE GRBs with each other, as well as with the predictions of this work may be explained by the fact that the prompt gamma-ray correlations used in the works of \citet{fenimore2000redshifts}, \citet{band2004gamma}, and \citet{yonetoku2004gamma} were constructed from a small sample of heterogeneously collected GRB events, and that the observed relations are likely severely affected by sample incompleteness. Despite the discrepancies in the redshift estimates of individual BATSE LGRBs, our predictions corroborate the findings of some previous studies \citep[e.g.,][]{ashcraft2007there, hakkila2009gamma} in that the probability of the existence of very high-redshift LGRBs in BATSE catalog is negligible. This is also evident from the dashed lines in the plot of Figure \ref{fig:histRedshift}, which represent the overall expected distribution of BATSE LGRBs according to the three LGRB rate densities considered.\newpar

     As illustrated in Figure \ref{fig:redshiftComparisonWithOldWorks}, the fact that the redshift estimates of \citet{yonetoku2004gamma} show the least disagreement with our predictions among all previous attempts, can be explained by noting the relative similarity of the assumptions in \citet{yonetoku2004gamma}, to infer the redshifts, with the findings of this work: \citet{yonetoku2004gamma} infer the redshifts based on the assumption of the existence of a tight positive correlation between the intrinsic spectral peak energy and the peak luminosity of LGRBs. Our modeling approach confirms the existence of such positive correlation (see Table \ref{tab:paraPostStat}), albeit with much higher dispersion and a different strength. In fact, the disparity in the predicted strength and regression-slope of this correlation can reasonably explain the non-negligible disagreement between our predictions and those of \citet{yonetoku2004gamma}.

    \begin{figure*}[tphb]
        \centering
        \begin{tabular}{ccc}
            \includegraphics[width=0.316\textwidth]{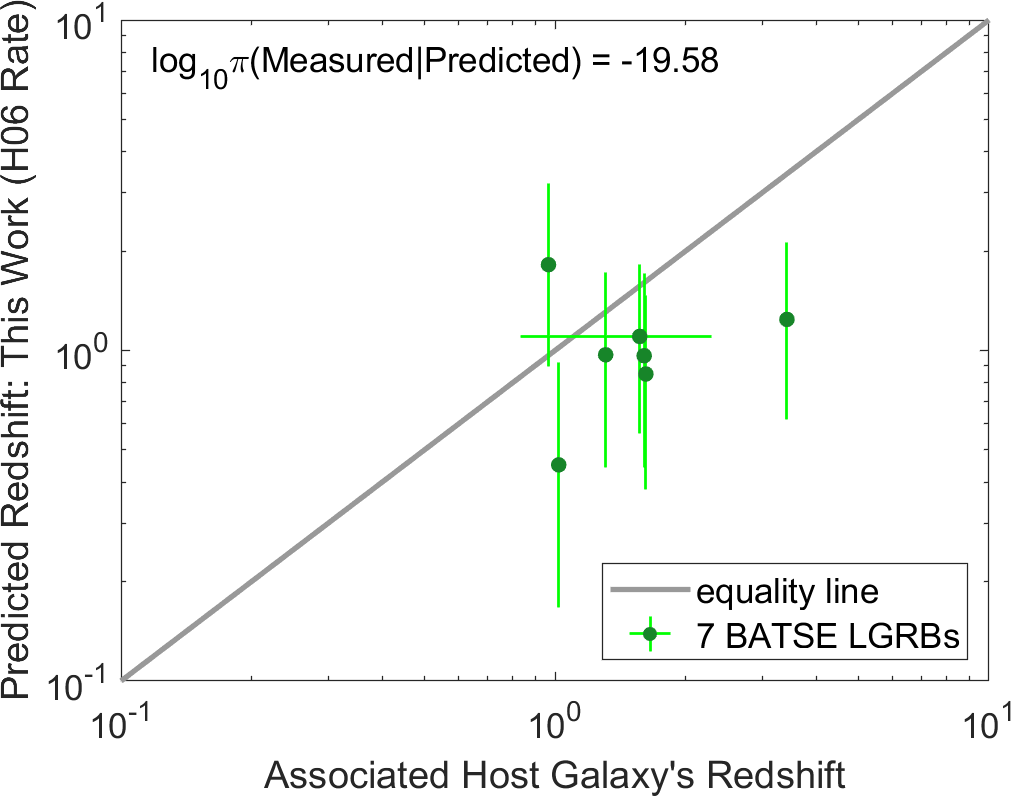} &
            \includegraphics[width=0.316\textwidth]{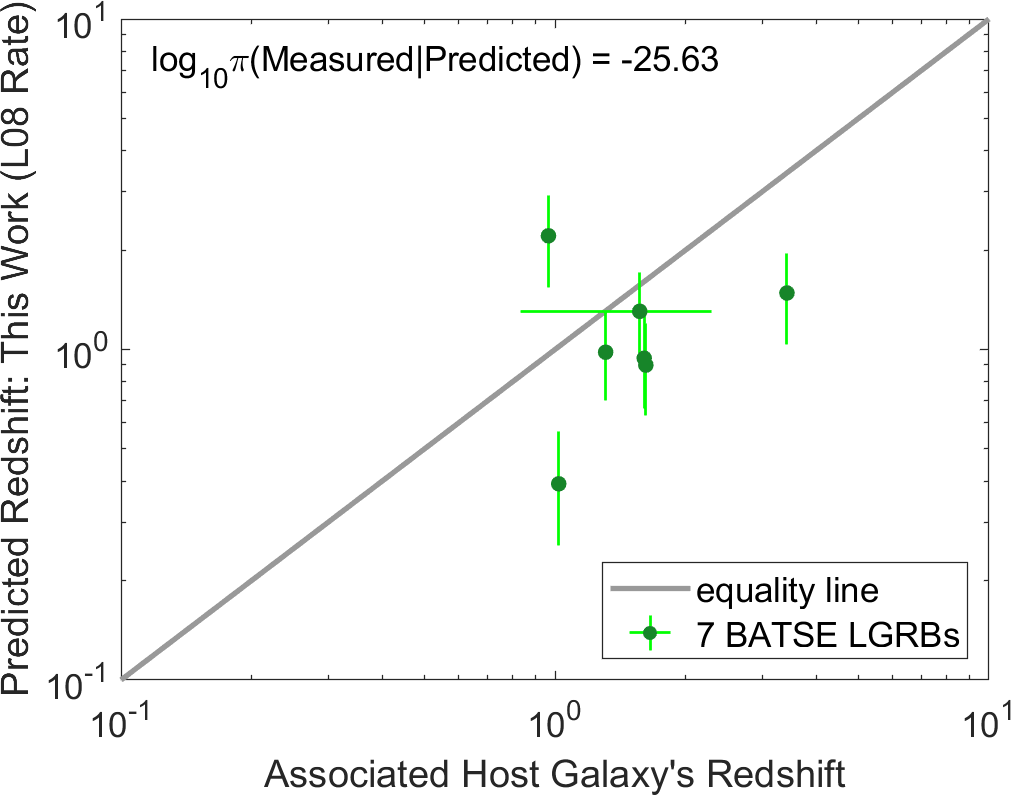} &
            \includegraphics[width=0.316\textwidth]{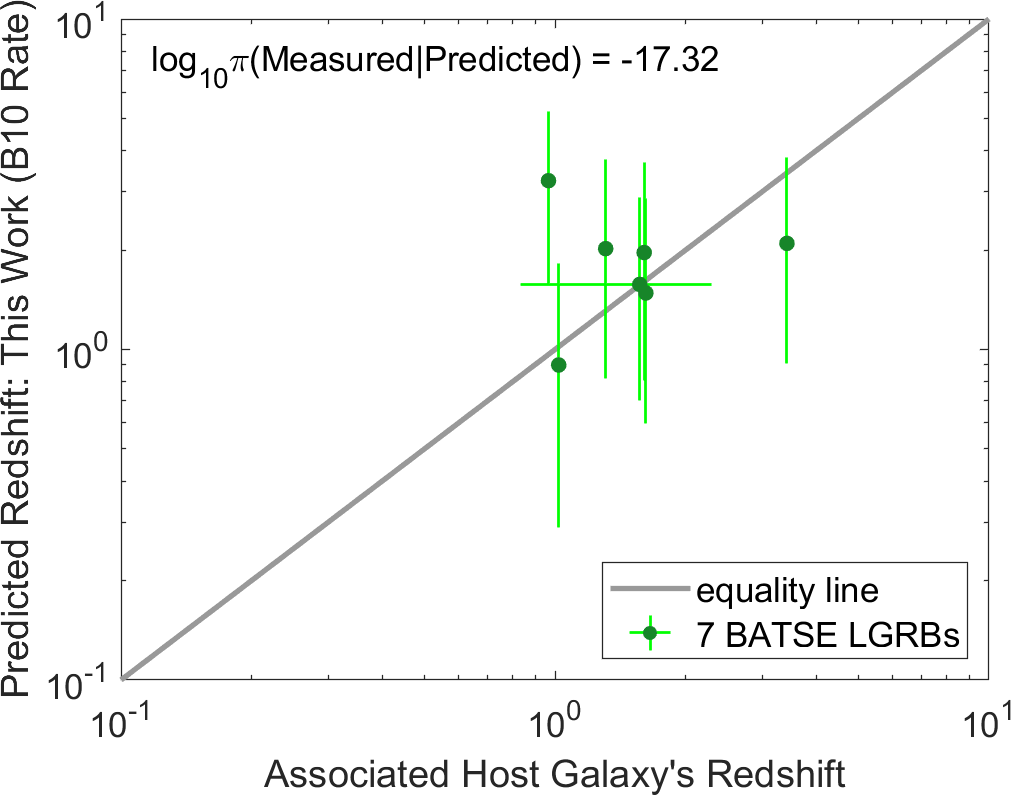} \\
        \end{tabular}
        \caption{A comparison of the predicted redshifts of a set of BATSE LGRBs with their reported measured redshifts in the literature, for the three LGRB rate density assumptions of \citetalias{hopkins2006normalization}, \citetalias{li2008star}, and \citetalias{butler2010cosmic}. The joint likelihood of the measured redshifts values representing the truth given the predicted redshifts are reported on each plot. The error bars represent the 90\% uncertainty intervals for the predicted redshifts and the 100\% uncertainty intervals for the measured redshifts (where available).\label{fig:knownRedshifts}}
    \end{figure*}

    Finally, we compare our predictions with the measured redshifts of BATSE events, where available. Out of 2704 BATSE LGRBs, only less than a dozen have measured redshifts through association with their potential host galaxies. In Figure \ref{fig:knownRedshifts}, we compare the reported redshifts of 7 such BATSE events (that also exist in our catalog) with their corresponding predicted redshifts in this work. Overall, the predicted redshifts for all three LGRB rate models are consistent with the host galaxy's redshifts within $90\%$ confidence level. Among the three, however, the predicted redshifts based on the LGRB rate of \citetalias{butler2010cosmic} show the highest level of consistency with the measured redshifts, followed by the predictions based on the LGRB rate of \citetalias{hopkins2006normalization}, followed by \citetalias{li2008star}. It is notable that at 50\% confidence, only the estimated redshifts based on the LGRB rate of \citetalias{butler2010cosmic} are consistent with the measured redshifts. In other words, more than 50\% (5 out of 7) of the measured redshifts fall within the corresponding predicted redshift range at 50\% confidence level for the LGRB rate of \citetalias{butler2010cosmic}, whereas this number for the two other LGRB rates is only 1 out of 7 BATSE events. This leads us to cautiously conclude that the LGRB formation rate may not exactly trace the star formation rate in the distant universe, corroborating the previous finding of \citetalias{butler2010cosmic}, \citet{shahmoradi2013multivariate}, and \citet{shahmoradi2015short}.\newpar

    Although very unlikely to be the case, one of the potential caveats of our presented redshift estimates is that, if an SGRB has been mistakenly classified as an LGRB in our catalog of 1366 by our classification method described in \ref{sec:methods:data}, then its estimated redshift may not be accurate. In addition, this work did not take into account the potential effects of the GRBs' jet beaming angle. In fact, a recent study by \citet{lazzati2013photospheric} finds that the different orientations of the GRB jet axis with respect to the observer could partially explain the observed LGRB brightness-hardness type relations. Such an effect will lead to more uncertainty in the predicted redshifts and perhaps could explain the lack of a complete perfect agreement between the known redshifts of a handful of BATSE LGRBs and their corresponding predicted redshifts in this work, as illustrated by the plots of Figure \ref{fig:knownRedshifts}.\newpar

\acknowledgments

This work would have not been accomplished without the vast time and effort spent by many scientists and engineers who designed, built and launched the Compton Gamma-Ray Observatory and were involved in the analysis of GRB data from BATSE Large Area Detectors.\newpar

\bibliographystyle{aasjournal}
\bibliography{./all}

\appendix

\begin{center}
\begin{longtable*}{cccccccccc}
        \caption[]{BATSE 1366 LGRB Redshift Estimates \label{tab:redshiftCatalog}} \\
        \hline \hline \\[0.5ex]
        Trigger &
        $\mu_{Z,H06}$ & PI$_{50\%}$ & PI$_{90\%}$ &
        $\mu_{Z,L08}$ & PI$_{50\%}$ & PI$_{90\%}$ &
        $\mu_{Z,B10}$ & PI$_{50\%}$ & PI$_{90\%}$
        \\[0.5ex] \hline \\[0.5ex]
        \endfirsthead
        %
        \multicolumn{10}{c}{{\tablename} \thetable{} -- Continued} \\[0.5ex]
        \hline \hline \\[0.5ex]
        Trigger &
        $\mu_{Z,H06}$ & PI$_{50\%}$ & PI$_{90\%}$ &
        $\mu_{Z,L08}$ & PI$_{50\%}$ & PI$_{90\%}$ &
        $\mu_{Z,B10}$ & PI$_{50\%}$ & PI$_{90\%}$
        \\[0.5ex] \hline \\[0.5ex]
        \endhead
        %
        \\[0.5ex] \hline \\[0.5ex]
        \multicolumn{10}{r}{{Continued on Next Page \ldots}} \\[0.5ex]
        \endfoot
        \\[0.5ex] \hline \hline
        \endlastfoot
        105 & 0.47 & $[0.31 , 0.59]$ & $[0.18 , 0.91]$ & 0.45 & $[0.38 , 0.51]$ & $[0.30 , 0.65]$ & 0.72 & $[0.44 , 0.92]$ & $[0.24 , 1.43]$ \\ [0.5ex]
107 & 1.42 & $[1.03 , 1.72]$ & $[0.71 , 2.43]$ & 2.05 & $[1.76 , 2.33]$ & $[1.35 , 2.78]$ & 1.92 & $[1.32 , 2.37]$ & $[0.86 , 3.49]$ \\ [0.5ex]
109 & 1.60 & $[1.14 , 1.92]$ & $[0.78 , 2.83]$ & 1.88 & $[1.65 , 2.09]$ & $[1.32 , 2.51]$ & 3.05 & $[2.22 , 3.76]$ & $[1.40 , 5.00]$ \\ [0.5ex]
110 & 2.16 & $[1.55 , 2.66]$ & $[1.02 , 3.69]$ & 3.07 & $[2.71 , 3.43]$ & $[2.14 , 3.96]$ & 3.20 & $[2.40 , 3.89]$ & $[1.57 , 5.13]$ \\ [0.5ex]
111 & 2.20 & $[1.56 , 2.71]$ & $[1.01 , 3.80]$ & 2.90 & $[2.58 , 3.23]$ & $[2.05 , 3.76]$ & 3.66 & $[2.79 , 4.33]$ & $[1.82 , 5.90]$ \\ [0.5ex]
114 & 1.82 & $[1.28 , 2.22]$ & $[0.86 , 3.19]$ & 2.47 & $[2.15 , 2.78]$ & $[1.69 , 3.31]$ & 3.00 & $[2.18 , 3.69]$ & $[1.37 , 4.94]$ \\ [0.5ex]
121 & 2.07 & $[1.48 , 2.53]$ & $[0.97 , 3.61]$ & 2.65 & $[2.34 , 2.93]$ & $[1.86 , 3.47]$ & 3.52 & $[2.67 , 4.19]$ & $[1.75 , 5.66]$ \\ [0.5ex]
130 & 1.52 & $[1.09 , 1.82]$ & $[0.75 , 2.67]$ & 1.82 & $[1.59 , 2.01]$ & $[1.27 , 2.40]$ & 2.85 & $[2.04 , 3.55]$ & $[1.28 , 4.73]$ \\
\end{longtable*}
\end{center}
{Notes: The column denoted by `Trigger' contains the trigger IDs of BATSE LGRBs. The columns denoted by $\mu_{H06}$, $\mu_{L08}$, $\mu_{B10}$ contain the predicted mean redshifts of individual BATSE LGRBs, based on the three LGRB rate model assumptions (\citetalias{hopkins2006normalization}, \citetalias{li2008star}, \citetalias{butler2010cosmic}) considered in this work as given by \eqref{eq:mz} and \eqref{eq:pz}. The rest of the columns denoted by `PI$_{50\%}$' and `PI$_{90\%}$' contain the lower and upper bounds of the 50\% and 90\% Prediction Intervals (i.e., the most probable ranges) of redshifts of individual BATSE LGRBs. This table as well as the full probability density functions of the redshifts of individual BATSE LGRBs are available for download at \url{https://github.com/shahmoradi/BatseRedshiftEstimates}.}

\newpage
\begin{center}
\begin{longtable*}{cccccc}
        \caption[]{The Spectral and Temporal Properties of BATSE 1366 LGRBs Used for Model Calibration\label{tab:BatseData}} \\
        \hline \hline \\[0.5ex]
        Trigger & $\pbol$ [ergs/s/cm$^2$] & $\sbol$ [ergs/cm$^2$] & $\epk$ [keV] & $\dur$ [s] & $\pph$ [photons/s/cm$^2$] \\[0.5ex] \hline \\[0.5ex]
        \endfirsthead
        %
        \multicolumn{6}{c}{{\tablename} \thetable{} -- Continued} \\[0.5ex]
        \hline \hline \\[0.5ex]
        Trigger & $\pbol$ [ergs/s/cm$^2$] & $\sbol$ [ergs/cm$^2$] & $\epk$ [keV] & $\dur$ [s] & $\pph$ [photons/s/cm$^2$] \\[0.5ex] \hline \\[0.5ex]
        \endhead
        %
        \\[0.5ex] \hline \\[0.5ex]
        \multicolumn{6}{r}{{Continued on Next Page \ldots}} \\[0.5ex]
        \endfoot
        \\[0.5ex] \hline \hline
        \endlastfoot
        105 & $4.778E-06$ & $7.362E-06$ & $9.078E+01$ & $5.184E+00$ & $9.938E+00$ \\ [0.5ex]
107 & $9.661E-08$ & $3.502E-07$ & $8.147E+01$ & $2.086E+02$ & $2.000E-01$ \\ [0.5ex]
109 & $1.822E-06$ & $8.063E-05$ & $1.995E+02$ & $9.018E+01$ & $3.350E+00$ \\ [0.5ex]
110 & $1.816E-07$ & $4.243E-06$ & $1.346E+02$ & $4.300E+02$ & $3.670E-01$ \\ [0.5ex]
111 & $3.067E-07$ & $7.492E-06$ & $5.236E+01$ & $9.818E+01$ & $5.890E-01$ \\ [0.5ex]
114 & $2.833E-07$ & $3.220E-06$ & $4.365E+01$ & $3.812E+02$ & $5.159E-01$ \\ [0.5ex]
121 & $5.572E-07$ & $2.359E-05$ & $1.334E+02$ & $7.501E+01$ & $1.128E+00$ \\ [0.5ex]
130 & $1.512E-06$ & $5.295E-05$ & $1.914E+02$ & $6.195E+01$ & $2.816E+00$ \\
\end{longtable*}
\end{center}
{Notes: The column denoted by `Trigger' contains the trigger IDs of BATSE LGRBs. The columns denoted by `$\pbol$' and `$\sbol$' contain the computed bolometric 1024 [ms] observed peak energy flux and total observed energy emission from the events in the energy range 100eV-20MeV with units of [ergs/s/cm$^2$] and [ergs/cm$^2$] respectively. The column denoted by `$\epk$' contains estimates of the observed spectral peak energies of BATSE LGRBs in units of [keV], extracted from \citet{shahmoradi2010hardness}. The column denoted by `$\dur$' contains the duration of BATSE LGRBs in seconds, [s], as measured by the time interval during which $90\%$ of the total observed energy emission is received. The column denoted by `$\pph$' contains the peak photon flux of BATSE events that is received in 1024 [ms] in BATSE's nominal detection energy window: 50-300 [keV], in units of [photons/s/cm$^2$]. This table is available for download at \url{https://github.com/shahmoradi/BatseRedshiftEstimates} in machine-readable format.}

\end{document}